\documentclass[12pt,a4paper]{article}
\usepackage{amsmath}
\usepackage[psamsfonts]{amssymb}
\numberwithin{equation}{section}

\newcommand{\br}{\mathbb R}
\newcommand{\bz}{\mathbb Z}
\newcommand{\bc}{\mathbb C}

\newcommand{\ket}[1]{{|#1\rangle}{}}

\newcommand{\nn}{\nonumber\\}
\DeclareMathOperator*{\Tr}{{\rm Tr}}
\DeclareMathOperator*{\re}{{\rm Re}}

\begin{document}
\begin{titlepage}

 \renewcommand{\thefootnote}{\fnsymbol{footnote}}
\begin{flushright}
 \begin{tabular}{l}
 SNUST-040301\\
  hep-th/0403081\\
 \end{tabular}
\end{flushright}

 \vfill
 \begin{center}
 \font\titlerm=cmr10 scaled\magstep4
 \font\titlei=cmmi10 scaled\magstep4
 \font\titleis=cmmi7 scaled\magstep4
 \centerline{\titlerm String Theory on Lorentzian {\titlei AdS}{\titleis 3} in
  Minisuperspace}
 \vskip 2.5 truecm

\noindent{ \large Yasuaki Hikida}\footnote{
E-mail: hikida@phya.snu.phys.ac.kr}
\bigskip

 \vskip .6 truecm
\centerline{\it School of Physics \& BK-21 Physics Division}
\vspace*{0.25cm}
\centerline{\it Seoul National University, Seoul 151-747 \rm KOREA}
 \vskip .4 truecm

 \end{center}

 \vfill
\vskip 0.5 truecm

\begin{abstract}
We investigate string theory on Lorentzian $AdS_3$ in the minisuperspace
approximation. The minisuperspace model reduces to the 
worldline theory of a scalar particle in the Lorentzian $AdS_3$.
The Hilbert space consists of normalizable wave functions, and we see
that the unitarity of the theory (or the self-adjointness of the
Hamiltonian) restricts the possible sets of wave functions.
The restricted wave functions have the property of probability
conservation (or current conservation) across the horizons. 
Two and three point functions are also computed.
In the Euclidean model
functional forms of these quantities are restricted by the $SL(2,\br)$
symmetry almost uniquely, however, in the Lorentzian model there are
several ambiguities left. 
The ambiguities are fixed by the direct computation of
overlaps of wave functions.
\end{abstract}
\vfill
\vskip 0.5 truecm

\setcounter{footnote}{0}
\renewcommand{\thefootnote}{\arabic{footnote}}
\end{titlepage}

\newpage


\section{Introduction}
\label{Intoduction}

Superstring theory on Anti-de Sitter (AdS) space attracts
many attentions recently due to the AdS/CFT correspondence 
\cite{AdSCFT,GKP,Witten}.
According to the AdS/CFT correspondence, superstring theory on 
$(d+1)$-dimensional Anti-de Sitter space is dual to $d$-dimensional
conformal field theory (CFT) defined at the boundary of AdS. 
This implies that correlators in AdS and those in CFT are
mappable each other.
However, superstring theory on AdS is difficult to deal
with, because we do not know how to quantize superstrings on a
target space including RR-flux in general (except for the pp-wave case
\cite{pp-wave,pp-wave2}).

The only tractable case is superstrings on $AdS_3$ with NSNS-flux,
which is dual to 2-dimensional CFT at the boundary
\cite{GKS,dBORT,KS}. Superstring theory on Euclidean $AdS_3$
can be described by $SL(2,\bc)/SU(2)$ Wess-Zumino-Novikov-Witten 
(WZNW) model. The WZNW model was investigated in
detail, and in particular the minisuperspace limit (the limit where we
neglect the dependence of the spatial coordinate $\sigma$) was discussed
in \cite{Gawedzki,Teschner1}, and three point function was obtained
exactly in \cite{Teschner2}. Superstring theory on {\it Lorentzian} 
$AdS_3$ can be described by $SL(2,\br)$ WZNW model, however this model
is much more difficult than its Euclidean counterpart (the difficulty
was discussed, for example, in \cite{Modular}). In particular,
the precise spectrum was discovered only recently \cite{MO1,MO2}.
Moreover, correlation functions in the $SL(2,\br)$ WZNW theory were
conjectured to be given by the analytic continuation of those in
$SL(2,\bc)/SU(2)$ WZNW model, and there is no direct derivation of them
(see, for the previous discussion, \cite{HS,Satoh,MO3}).

In this paper, we would like to deal with the Lorentzian model directly
although in the minisuperspace limit.
As it stands now, complete definition of string theory on a curved
Lorentzian spacetime is not known. 
The only proposal is to utilize the analytic continuation of the
Euclidean counterpart.
In the minisuperspace limit, one can do better as the theory is reduced
to a quantum mechanical system, and hence can treat the theory directly
in the Lorentzian signature. 
The Hilbert space is constructed by square integrable wave
functions, however the wave functions are not compatible to the
unitarity of the theory in general. We follow a general theory of
self-adjointness of a linear operator (for a review, see
\cite{Selfadjoint}) to construct the domain of self-adjoint
Hamiltonian which includes a set of restricted wave functions.

A situation analogous to this is timelike Liouville theory
\cite{S,GS2}. Timelike Liouville theory arises as a continuum
worldsheet description of spacelike S-brane \cite{GS1} and
rolling tachyon \cite{Sen}. In \cite{S,GS2,MSY,ST,mini}, timelike
Liouville theory was defined as the analytic continuation of
spacelike Liouville theory, and correlation functions are calculated.
Recently, in \cite{mini2}, utilizing the reduced model of  $SL(2,\br)$ WZNW
model \cite{reduced1,reduced2}, the minisuperspace limit of timelike
Liouville theory was investigated directly, and the minisuperspace model
was compared with the analytic continued one.

This paper is organized as follows.
In section \ref{momentum}, we investigate normalizable states of the
minisuperspace $AdS_3$ string theory. 
The self-adjointness condition is equivalent to the boundary condition
of wave functions, and we construct a one parameter family of
self-adjoint extensions of the Hamiltonian.
In the global coordinates of Lorentzian AdS, the normalizable wave
functions are unique as in \cite{Balas1}, and there is unique
self-adjoint extension. However, in the Poincar\'e coordinates adopted in
the context, the normalizable wave functions are 
not unique, and the self-adjointness condition restricts the wave
functions to those satisfying the boundary condition. There are horizons
in the Poincar\'e coordinates, and the boundary condition means the
probability conservation (or the current conservation) across the
horizons.

In section \ref{coordinate}, we construct primary fields and compute
correlation functions.
In order to construct primary fields, we introduce parameters
$(x,\bar x)$, which may be interpreted as coordinates of the boundary
of $AdS_3$ in the sense of the AdS/CFT correspondence \cite{dBORT,KS}. 
The advantage of introducing the parameters is
that the transformation of the $SL(2,\br)$ symmetry can be generated by
differential operators, and primary fields are constructed as solutions
to differential equations. We can find that the primary field is
precisely the Fourier transform of wave function constructed in section
\ref{momentum}. The solution is not unique and it corresponds to the
non-uniqueness of the wave functions. Two or three point functions
are also obtained as solutions to differential equations. 
The solutions are almost unique in the Euclidean theory, however the
solutions in the Lorentzian theory have more ambiguities undetermined by
the $SL(2,\br)$
symmetry. In the minisuperspace limit, we can compute the correlation
functions as overlaps of the wave functions and fix the ambiguities, but
in the full CFT case, we may have to use the symmetry to compute the
correlation functions as in the case of the Euclidean theory
\cite{Teschner2}.

Section \ref{conclusion} is devoted to conclusion and discussions. 
In appendix \ref{formulae}, we summarize various formulae relevant for
computations in this paper.


\section{Spectrum of the minisuperspace model}
\label{momentum}

String theory on the Euclidean $AdS_3$, which is known to be described by 
$SL(2;\bc)/SU(2)$ WZNW model, has been investigated for a decade,
for example, in \cite{Gawedzki,Teschner1,Teschner2}.
The theory is an example of non-rational conformal field theory, and in
general non-rational conformal field theory is difficult to analyze
because it has infinitely many primary fields; only spacelike Liouville
theory and $SL(2;\bc)/SU(2)$ WZNW model were solved.
The truncation in the minisuperspace limit (in case of spacelike
Liouville theory, see \cite{Seiberg}) gives a theory only with zero-mode
subspace, however it still includes infinite dimensional primary
fields. The truncations in spacelike Liouville theory and
$SL(2;\bc)/SU(2)$ WZNW model 
played important roles on the investigation of full theories, so we
could expect that the minisuperspace limit of $SL(2;\br)$ WZNW model
gives insights into the understanding of the full CFT model. 
In this section, we will investigate the spectrum by canonical
quantization of the minisuperspace model.


\subsection{$AdS_3$ space and $SL(2;\br)$ group elements}
\label{AdS3}

The Lorentzian $AdS_3$ space is defined as a hypersurface
\begin{equation}
 - X_0^2 + X_1^2 + X_2^2 - X_3^2 = - L_{AdS}^2 ~,
\end{equation}
in $(2+2)$-dimensional embedding flat space $\br^{2,2}$ of signature
$(-,+,+,-)$. 
It is convenient to represent the hypersurface in terms of
matrix $g$:
\begin{equation}
  g = \frac{1}{L_{AdS}}
\begin{pmatrix}
X_1 + X_3 & X_0 - X_2 \\
- X_0 - X_2 & - X_1 + X_3
\end{pmatrix}  \qquad \mbox{where} \qquad \det g = 1 ~,
\end{equation}
viz. group elements of $SL(2, \br)$. The isometry
$SO(2,2) \simeq SL(2,\br)_L \times SL(2,\br)_R$
on the Lorentzian $AdS_3$ is then realized as left and right action on
the group element $g$. The conformal field theory whose target space is
the $SL(2,\br)$ group manifold, viz. the $SL(2,\br)$ WZNW model, then
describes string propagation on the Lorentzian $AdS_3$.

One useful parametrization of the group elements is
\begin{equation}
 g = e^{i \theta_L \sigma_2} e^{\rho \sigma_3} e^{i \theta_R \sigma_2}~,
 \qquad \theta_L = \frac12(t + \varphi) ~, 
 \quad \theta_R = \frac12(t - \varphi) ~,
\end{equation}
where $\sigma_i$ $(i=1,2,3)$ represents the Pauli matrices.
These parameters correspond to the global coordinates of $AdS_3$, 
and the metric is written as
\begin{equation}
 d s^2 = - \cosh ^2 \rho d t^2 + d \rho^2 
      + \sinh ^2 \rho d \varphi ^2 ~.
\end{equation}
The parameters $\rho$ and $\varphi$ run $- \infty < \rho < \infty$ and 
$0 \leq \varphi < 2 \pi$, respectively. The boundaries of $AdS_3$ at 
$\rho=\pm \infty$ are connected to each other, so there is only one
boundary.
For the single cover of $AdS_3$ space, the time direction is periodic 
$0 \leq t < 2 \pi$ and covers the whole spacetime once. 
For the universal cover of $AdS_3$ space, we unwrap the closed timelike
curve $-\infty < t < \infty$. We will consider only the universal cover
of $AdS_3$ without mention. The Euclidean $AdS_3$ can be obtained from the
Lorentzian $AdS_3$ by the Wick rotation $X_0 \to i X_E$ or equivalently
$t \to i t$.

There is another useful parametrization of the
group elements $g$, based on the Gauss decomposition:
\begin{equation}
 g =
\begin{pmatrix}
1 & \bar{\gamma}\\
0 & 1
\end{pmatrix}
\begin{pmatrix}
z & 0\\
0 & 1/z
\end{pmatrix}
\begin{pmatrix}
1 & 0\\
\gamma & 1
\end{pmatrix} ~.
\label{para}
\end{equation}
In this parametrization, the Lorentzian $AdS_3$ is described by so-called
Poincar\'e coordinates
\begin{eqnarray}
d s^2 = {1 \over z^2} \left[ d z^2
+ d \gamma d \bar{\gamma} \right] ~.
\end{eqnarray}
There are two patches for the single cover of $AdS_3$ space, and
the coordinate $z$ ranges over $(-\infty,-0)$ and $(+0, + \infty)$
for the each patch. The other coordinates $\gamma, \bar \gamma$ range
over $-\infty < \gamma,\bar\gamma < \infty$; they are independent and
should not be thought as complex conjugates each other. 
The patch with $z>0$ covers half of the spacetime $- X_1 + X_3 > 0$
and there is a boundary at $z=+0$ and a horizon at $z=+\infty$. 
The other half of the spacetime $- X_1 + X_3 < 0$ is obtained by
replacing $z \leftrightarrow - z$, and the patch is glued at the horizon
$z= - \infty$ with the other patch.
In order to cover whole the universal cover of $AdS_3$, we need
infinitely many Poincar\'e patches and glue the patches with $z>0$ and $z<0$
alternately. 
If we only deal with a patch, then we often adopt the coordinate 
$z = \pm e^{-\phi}$ $-\infty < \phi < + \infty$ with the metric
\begin{equation}
 d s^2 =  d \phi^2 + e^{2\phi} d \gamma d \bar{\gamma} ~.
\end{equation}
Note that the Euclidean $AdS_3$ is obtainable by the Wick rotation:
$X_0 \to i X_E$, viz. $t' \to i t'$ in the parametrization of
$\gamma = 1/2(\theta + t')$ and $\bar{\gamma}=1/2(\theta - t')$. Then,
in the Euclidean $AdS_3$, $\gamma$ and $\bar \gamma$ are complex conjugates
each other.

\subsection{Minisuperspace limit of $SL(2;\br)$ WZNW model}

String dynamics on Lorentzian $AdS_3$ is described by the $SL(2,\br)$
WZNW model action
\begin{equation}
 S = \frac{k}{4 \pi} \int_\Sigma d \tau d \sigma
  \Tr [\partial g^{-1} \bar \partial g]
+ k \Gamma_{WZ}~,
\end{equation}
where $\Gamma_{WZ}$ refers to the Wess-Zumino term.
We take the Lorentzian worldsheet of topology 
$\br \times \mathbb{S}^1$, and denote derivatives as 
$\partial= 1/2(\partial_\tau + \partial_\sigma)$ and
$\bar{\partial}= 1/2(\partial_\tau - \partial_\sigma)$.
As a parametrization of the group elements, we will adopt the Gauss
decomposition (\ref{para}) even though we need infinitely many patches.
As we will see below, the minisuperspace model reduces to a quantum
mechanics, and the states can be represented as wave functions.
The form of wave functions is simpler in the Poincar\'e coordinates than in
the global coordinates.\footnote{In the global coordinates, the wave
functions are expressed by hypergeometric functions. Later we compute
overlaps of the wave functions, and it is more complicated to perform
the integrals of three hypergeometric functions.}
The cost of this is that we have to take care of the connection between
each two adjacent patches.
Of course, it is just a technical problem and the physics must be the
same in the both coordinate systems.

In the chosen Gauss decomposition (\ref{para}), string dynamics on
Lorentzian $AdS_3$ is described by the $SL(2,\br)$ WZNW model
action
\begin{equation}
 S = \frac{k}{2 \pi} \int_\Sigma d \tau d \sigma
   {1 \over 2 z^2} [ 2 \partial z \bar{\partial} z + 
   \partial \bar{\gamma} \bar{\partial} \gamma + 
   \partial \gamma \bar{\partial} \bar \gamma] + k \Gamma_{WZ} ~.
\end{equation}
As mentioned above, there are two types of patches, and one of them has
the coordinate $z>0$ and the other has $z<0$. We shall adopt
$\arg z = \pi$ in the latter type of patch for definiteness.
In the minisuperspace limit, the string is treated as a rigid body, so
the worldsheet fields $(z,\gamma,\bar \gamma)$ become
independent of the $\mathbb{S}^1$ coordinate $\sigma$. In this limit,
the Wess-Zumino term drops out automatically, and the action is reduced to
\begin{equation}
 S = \frac{k}{4} \int d \tau
{1 \over z^2} [ (\partial_\tau z)^2 +
\partial_\tau \bar \gamma \partial_\tau \gamma]
= \frac{k}{4} \int d \tau [(\partial_\tau \phi)^2 + e^{2 \phi}
\partial_\tau \overline \gamma \partial_\tau \gamma]~.
\label{miniaction}
\end{equation}
The classical Hamiltonian can be calculated as\footnote{We rescale the
overall factor $k \to 1$ because it affects only the overall normalization.}
\begin{equation}
 H_{ws} = (p_{\phi})^2 + 4 e^{-2\phi}p_{\gamma} p_{\bar\gamma} ~.
\end{equation}
In general, for a given classical Hamiltonian, corresponding quantum 
Hamiltonian is afflicted by operator--ordering ambiguity. 
Upon quantization, the canonical momentum $p_\phi$ conjugate to
the radial coordinate $\phi$ is promoted to
$p_\phi = z \partial_z$, so the quantum Hamiltonian would take
one of the following forms:
\begin{eqnarray}
 H_{ws} = z^a \frac{\partial}{\partial z} z^b \frac{\partial}{\partial z} z^c
 +  4 z^2 \frac{\partial}{\partial \gamma} \frac{\partial}{\partial
\bar \gamma} \quad\quad \mbox{where} \quad\quad   a+b+c=2 ~.
\label{QH}
\end{eqnarray}
A direct way of prescribing the quantum Hamiltonian is by 
taking the invariant Lichnerowicz operator on $SL(2;\br)$ group
manifold: 
\begin{equation}
H_{ws} := \Box_{\rm AdS} \equiv 
 {1 \over \sqrt{- \det g}}\partial_m 
 \left(\sqrt{- \det g} g^{mn}\partial_n\right) ~. 
\label{H_ws}
\end{equation}
It is readily seen that, in this prescription, the
momentum operators are Weyl-ordered.
In fact, the eigenfunctions of the quantum Hamiltonian can be
interpreted as wave functions satisfying a Klein-Gordon equation in the
Lorentzian $AdS_3$
\begin{equation}
 \square_{\rm AdS} \Phi (z,\gamma,\bar \gamma) =
   4 j (j-1) \Phi(z,\gamma,\bar \gamma) ~, \qquad
 \square_{\rm AdS} \equiv z^2 \frac{\partial ^2}{\partial z^2}
           - z \frac{\partial}{\partial z} + 4 z^2
 \frac{\partial}{\partial \gamma} \frac{\partial}{\partial \bar \gamma} ~.
\label{KG}
\end{equation}
Here the mass square is written in terms of the Casimir invariant of the
$SL(2,\br)$ Lie algebra $(c_2 = j(j-1))$ for later convenience.

\subsection{Normalizable wave functions}

The wave function satisfying the Klein-Gordon equation (\ref{KG}),
or more precisely, the eigenfunction of the quantum Hamiltonian,
is given by reduction of the phase-space:
\begin{eqnarray}
 \Phi (z,\gamma,\bar \gamma)=
    e^{i \lambda \gamma + i \mu \bar \gamma } z \, U(z)~,
\label{decomp}
\end{eqnarray}
where $U(z)$ is the reduced wave function, referred as Liouville
wave function, obeying the zero-energy Schr\"odinger equation:
\begin{equation}
 \left(z^2 \frac{\partial ^2}{\partial z^2}
      + z \frac{\partial}{\partial z} -V(z) \right) \, U(z) = 0
\qquad \mbox{where} \qquad V(z) = 4 z^2 \lambda \mu
      - (2j-1)^2 ~ .
\label{KG2}
\end{equation}
The Liouville wave function $U(z)$ is solved in
general by a linear combination of the Bessel functions. Depending
on the reduction branches, two distinct behaviors are expected.
For $\lambda \mu > 0$, the `potential' $V(z)$ is bounded from
below. For $\lambda \mu < 0$, the potential is not bounded from
below, so a care should be exercised in this case
by prescribing carefully behavior of the wave functions at the
``boundary''.

We will look for the solutions to (\ref{KG2})
which are square normalizable (including delta functional normalizable)
with respect to the inner product 
\begin{equation}
 \langle \Phi_2 , \Phi_1 \rangle
   \equiv \int_{SL(2,\br)} d g \, \Phi_2^* (g) \Phi_1 (g) ~,
\label{innerprod_p}
\end{equation}
where $d g \equiv d \gamma d \bar \gamma d z |z|^{-3}$ is the
$SL(2,\br)$-invariant measure.
As mentioned above, there are infinitely many pairs of two adjacent
patches; one patch has the coordinate $z>0$ and the other has $z<0$.
We assume that the wave functions on the each pair of patches are
the same, and for this reason we will pick up one of the pairs.\footnote{
This might be justified by the fact that physics in the global
coordinates and in the Poincar\'e coordinates must be same. There is
time-translation invariance in the global coordinates, therefore it is
natural to assume that wave functions in next pair of Poincar\'e
patches are the same as those in the original pair.} Then, the inner
product we will use is written in
\begin{eqnarray}
 &&\langle \Phi_2 , \Phi_1 \rangle =  
 \langle \Phi_2 , \Phi_1 \rangle_- +  \langle \Phi_2 , \Phi_1 \rangle_+ 
  \label{innerprod}\\
   &&\qquad = \int^{\infty}_{-\infty} d\gamma \int^{\infty}_{-\infty} 
         d\bar \gamma \int^{-0}_{-\infty} \frac{dz}{|z|^3} 
   \Phi_2^* (g) \Phi_1 (g) 
   + \int^{\infty}_{-\infty} d\gamma \int^{\infty}_{-\infty} 
         d\bar \gamma \int^{\infty}_{+0} \frac{dz}{|z|^3} 
   \Phi_2^* (g) \Phi_1 (g) ~. \nonumber
\end{eqnarray}
In the following, the square integrability is examined with respect to
the above inner product.

For $z>0$, the square integrable solutions to the Klein-Gordon equation
(\ref{KG2}) with real valued eigenvalues $4j(j-1) \in \br$ are
given by 
 \begin{align}
  &a_1 K_{2j-1} \left( 2\sqrt{\lambda \mu} z\right)  
  &(\lambda \mu > 0 ,j = 1/2 + i \br) \label{K} ~,\\
  &a_2 J_{2j-1} \left( 2\sqrt{- \lambda \mu} z\right) &
 (\lambda \mu < 0 , j > 1/2) \label{DJ} ~,\\
  &a_3 J_{2j-1} \left( 2\sqrt{- \lambda \mu} z\right) +
  a_4 J_{1-2j} \left( 2\sqrt{- \lambda \mu} z\right) &
 (\lambda \mu < 0 ,j = 1/2 + i \br) ~,\label{CJ}
 \end{align}
with $z$-independent constants $a_{i} \in \bc, ~ i=1,2,3,4$
(which may have dependence on $j$). 
We denoted $J_{2j-1} (x)$ as the Bessel
functions of the first kind and $K_{2j-1} (x)$ as the modified Bessel
function of the second kind, respectively. 
Depending on the sign of $\lambda \mu$, the potential $V(z)$ in
(\ref{KG2}) pushes the wave function either to $z = 0$ or
$\infty$, so an appropriate `boundary' condition needs to be
prescribed at $z = 0, \infty$ so that the Hamiltonian (\ref{H_ws})
maintains self-adjointness.\footnote{See, for example,
\cite{Selfadjoint} for self-adjointness of Sturm-Liouville operators.} 
For $z>0$, the solutions are
\begin{align}
  &b_1 K_{2j-1} \left( 2\sqrt{\lambda \mu} z\right)  
  &(\lambda \mu > 0 ,j = 1/2 + i \br) \label{K2} ~,\\
  &b_2 J_{2j-1} \left( 2\sqrt{- \lambda \mu} z\right) &
 (\lambda \mu < 0 , j > 1/2) \label{DJ2} ~,\\
  &b_3 J_{2j-1} \left( 2\sqrt{- \lambda \mu} z\right) +
  b_4 J_{1-2j} \left( 2\sqrt{- \lambda \mu} z\right) &
 (\lambda \mu < 0 ,j = 1/2 + i \br) ~,\label{CJ2}
\end{align}
with  $z$-independent constants $b_{i} \in \bc, ~ i=1,2,3,4$. 
If there is no interaction between patches, then we can choose the
coefficients $b_i$ independent to $a_i$. Otherwise, $b_i$ depend on $a_i$.

\subsection{Self-adjointness of the Hamiltonian}

In examining self-adjointness of the Hamiltonian, 
we will first see an abstract theory on how to make an operator
self-adjoint. 
Let us denote ${\cal H}$ as a Hilbert space and $A$ as an operator
acting on a domain $D(A) \subset {\cal H}$. 
In our case, Hilbert space  ${\cal H}$ is made from the square
integrable wave functions, and linear operator is given by the quantum
Hamiltonian $H_{ws}$ (\ref{H_ws}).
Adjoint operator $A^*$ is defined by $\Psi ' = A^* \Psi$
for $\Psi \in D(A^*) \subset {\cal H}$ satisfying
\begin{equation}
 \langle \Psi , A \Phi \rangle = \langle \Psi ' , \Phi \rangle ~,\qquad
  {}^\forall \Phi \in D(A) ~,
\end{equation}
and the operator $A$ is called symmetric if
\begin{equation}
 \langle \Psi , A \Phi \rangle = \langle A \Psi ,\Phi \rangle ~,\qquad
  {}^\forall \Phi \in D(A)~,\quad {}^\forall \Psi \in D(A) \subset D(A^*) ~.
\end{equation}
In particular, the operator is self-adjoint if $D(A) = D(A^*)$.
In our case, we first construct the domain for $H_{ws}$ to be 
symmetric, and then we extend the operator to be
self-adjoint. An extension $B$ of an operator $A$ is defined by 
$D(B) \supset D (A)$ with $B=A$ on $D(A)$, and an extension $B$ of
a symmetric operator $A$ is self-adjoint if 
$D(A^*) \supset D(B^*) = D(B) \supset D (A)$.

We use a generic theory concerned with self-adjoint extension of a
symmetric operator (see, for example, \cite{Selfadjoint}).
For a symmetric operator $A$, we can decompose the domain $D(A^*)$ as
\begin{equation}
 \phi = \psi + \xi_+ + \xi_- ~, ~~
 \phi \in D(A^*) ~,~ \psi \in D(A) ~,~
 \xi_+ \in K_+ (A^*) ~,~ \xi_- \in K_- (A^*) ~, 
\end{equation}
where\footnote{The eigenvalues $\pm i$ could be
replaced with an arbitrary pair of complex numbers $c$, $c^*$ with 
${\rm Im} \, c \neq 0$.}
\begin{align}
 &K_+ (A^*) := {\rm Ker} (A^* - i) ~,&
 &K_- (A^*) := {\rm Ker} (A^* + i) ~.
\end{align}
Roughly speaking, if an operator $A$ is not self-adjoint, then there
are eigenfunctions of $A^*$ whose eigenvalues have imaginary part.
Then, we can see that there exists self-adjoint extensions if we can
construct extensions of the symmetric operator as
\begin{equation}
 D_{U}(B) = \{ \phi | \phi = \psi + \xi_+ + U \xi_+ , \psi \in D(A),
 \xi_+ \in K_+ (A^*), U \xi_+ \in K_- (A^*) \} ~
\label{domainU}
\end{equation}
with unitary transformation $U:\psi_+ \to \psi_-$. 
In other words, if deficiency indices $(d_+,d_-)$ defined by
\begin{align}
 &d_+ := \dim K_+ (A^*) ~,&
 &d_- := \dim K_- (A^*) 
\end{align}
are the same $d_+ = d_-$, then there exist a unitary transformation $U$
and a family of self-adjoint extensions parametrized by $U$.
In particular, in case of $d_+=d_-=0$, there is unique self-adjoint
extension $B = A$.

\subsection{Domain of the self-adjoint Hamiltonian}

Let us apply the abstract theory to our case.
The Hilbert space ${\cal H}$ consists of the square integrable
functions of $g \in SL(2,\br)$ with respect to the inner product
(\ref{innerprod}). For the following domain of the quantum Hamiltonian
$H_{ws}$ (\ref{H_ws})
\begin{equation}
 D(H_{ws}) = \{ \Phi | \Phi \in {\cal H}  , ~ H_{ws} \Phi \in {\cal H} , ~
         U(z)|_{z=\pm 0,\pm \infty} = 0 =  
         z \partial_z U(z)|_{z=\pm 0,\pm \infty}  \} ~,
\label{domain}
\end{equation}
the Hamiltonian is symmetric. This can be seen from that the symmetric
condition
\begin{equation}
 \langle H_{ws} \Phi_2 , \Phi_1 \rangle =  \langle \Phi_2 ,  
  H_{ws} \Phi_1 \rangle ~,\quad
 \Phi_1 \in D(H_{ws}) ~,~ \Phi_2 \in D (H_{ws}) \subset D (H_{ws}^*) 
\end{equation} 
can be rewritten as the form of boundary conditions
\begin{equation}
  \left. z \left(U^*_2 (z) \partial_z U_1 (z) -
  \partial_z  U^*_2 (z) U_1 (z) \right) 
    \right|^{ - 0}_{- \infty}
  -  \left. z \left(U^*_2 (z)\partial_z U_1 (z)-
  \partial_z  U^*_2 (z) U_1 (z) \right) 
    \right|_{+ 0}^{ + \infty} = 0~.
\label{BC}
\end{equation}
Since the boundary condition (\ref{BC}) is always satisfied for
${}^\forall \Phi_1 \in D(H_{ws})$, there is no need to assign boundary
conditions on $\Phi_2 \in D(H^*_{ws})$, so 
$D(H_{ws}) \subset D(H_{ws}^*)$.
As we will see below, the deficiency indices for  $\lambda \mu >0$ and
$\lambda \mu < 0$ cases are different, so
we investigate the self-adjointness of the symmetric Hamiltonian $H_{ws}$
(\ref{H_ws}) for the each case separately.

\subsubsection{$\lambda \mu >0$ case}

In this case, we can see that there are no eigenfunctions 
$H_{ws}\Psi = \pm i \Psi$ for $\Psi \in {\cal H}$, so
the deficiency indices are $(0,0)$. Therefore, the Hamiltonian with the
domain (\ref{domain}) is self-adjoint. 
The wave functions we have constructed (\ref{K}) and (\ref{K2}) are
included in the domain, so eigenfunctions of self-adjoint
Hamiltonian are given by (\ref{K}) and (\ref{K2}) with arbitrary $a_1$
and $b_1$. 
The each term in the boundary condition (\ref{BC}) vanishes by itself,
and there are no interactions between the patches (which is equivalent
to the independence of $a_1$ and $b_1$).
We denote these wave functions as $(j=1/2+i \omega,\omega \in \br)$
\begin{equation}
 \Phi_{\omega}^{\lambda , \mu} =
 \frac{\sqrt{2} (\lambda \mu)^{i \omega}}{\Gamma(2i \omega)}
 e^{i\lambda\gamma + i\mu\bar\gamma}
     z \, K_{2 i \omega} \left(2\sqrt{\lambda\mu} 
     e^{i \delta(z) }z\right) ~,\quad  
     \delta (z) = \frac{\pi}{2} ({\rm sgn} \ z - 1) ~.
\label{WK}
\end{equation}
We chose the normalization factors so that the two point function
is given by (see, e.g., \cite{ST1})
\begin{equation}
 \langle \Phi_{\omega_2 }^{\lambda_2 , \mu_2},
 \Phi_{\omega_1}^{\lambda_1 , \mu_1} \rangle
 = (2\pi)^3 \delta(\lambda_1 - \lambda_2) \delta(\mu_1 - \mu_2)
 \Big[
  \delta(\omega_1 - \omega_2)
 + R (\omega_1) \delta(\omega_1 + \omega_2) \Big] ~,
\label{2pt}
\end{equation}
where
\begin{equation}
R(\omega) = - (\lambda \mu)^{2i\omega}
 \frac{\Gamma(1-2i \omega)}{\Gamma(1+2i\omega)} ~.
\label{RA}
\end{equation}
Physical meaning of the factor $R(\omega)$ is extracted from
asymptotic behavior of the wave function near the boundary  $z \to \pm 0$, 
equivalently, $\phi \to \infty$ (where the Liouville potential
vanishes). The wave function becomes a linear combination of
incident and reflected plane waves:
\begin{equation}
 \Phi_{\omega}^{\lambda , \mu} \sim
 e^{i\lambda\gamma + i\mu\bar\gamma} e^{-\phi}
 \Big[ e^{2 i \omega \phi} + R(\omega)e^{- 2 i \omega \phi} \Big] ~.
\label{AB}
\end{equation}
It is evident that $R(\omega)$ denotes the reflection amplitude from
the Liouville potential.

\subsubsection{$\lambda \mu < 0$ case}

The eigenfunctions of $H_{ws}$ with eigenvalues $\pm i$ 
inside the Hilbert space ${\cal H}$ are given by
\begin{equation}
 \begin{split}
 U_{k_0}(z)&= 
 \begin{cases}
 d_1 J_{k_0} (2 \sqrt{- \lambda \mu}z) &{\rm for} ~ z>0 ~,\\
 e_1 e^{-k_0 \pi i}J_{k_0} (2 \sqrt{- \lambda \mu}z) &{\rm for} ~ z<0 ~,
 \end{cases}\\
 U_{k_0^*}(z)&= 
 \begin{cases}
 d_2 J_{k_0^*} (2 \sqrt{- \lambda \mu}z) &{\rm for} ~ z>0 ~,\\
 e_2 e^{-k_0^* \pi i} J_{k_0^*} (2 \sqrt{- \lambda \mu}z) &{\rm for} ~ z<0 ~,
 \end{cases} 
\label{k_0}
\end{split}
\end{equation}
with $k_0 = \sqrt{1+i}$. There are two independent eigenfunctions for
each eigenvalue $\pm i$, therefore the deficiency indices are $(2,2)$.
Since the unitary transformation $\hat M:U_{k_0} \to U_{k_0^*}$ has four
real parameters, 
there is four parameter family of self-adjoint
extensions:
\begin{equation}
 D_M (H_{ws}) = \{ \Phi | \Phi = \Psi + \Xi_M, 
    \Psi \in D(H_{ws}) \} ~,
\label{sfdomain}
\end{equation}
where we defined
\begin{align}
 \Xi_M &=  e^{i\lambda \gamma + i \mu \bar \gamma}  z \, U_M (z) ~,
  &U_M (z) &=  U_{k_0} (z) + \hat M  U_{k_0} (z) ~.
\label{theta}
\end{align}
The unitary transformation may be expressed by $2 \times 2$ unitary
matrix $M$ acting on the coefficients on (\ref{k_0}) as
\begin{align}
  \begin{pmatrix}
    d_2 \\ e_2
  \end{pmatrix}
   &= M 
 \begin{pmatrix}
    d_1 \\ e_1
  \end{pmatrix} ~,
 &M &= \exp \left(i \sum_{i=0}^{3} q_i \sigma_i\right) ~,
\label{theta2}
\end{align}
with $\sigma_0 \equiv {\rm diag} (1,1)$.

The self-adjointness parameters $q_i$ fix the boundary behaviors of
wave functions since the contribution from the $\Xi_M$ part in
(\ref{sfdomain}) is dominant near the boundary.
If we do not restrict the norms of the two eigenfunctions $U_{k_0}$ and 
$U_{k_0^*}$ (\ref{k_0}), then the domain includes rather arbitrary
wave functions because the general solutions to the second order
differential equations are given by linear combinations of two
independent solutions. The symmetric condition corresponds to the
boundary condition (\ref{BC}), which implies current conservation as
we will see below in more detail, therefore the restriction in
(\ref{sfdomain}) is equivalent to the restriction of the wave functions
to preserve the total current. There are non-trivial boundary
behaviors only in the $z \to \pm \infty$ region, therefore we could say
that the self-adjoint parameters determine how the current flows across
the horizons.

In order to see the relation between the self-adjoint parameters and the
boundary behaviors of wave functions, we set some of the parameters
zero. Let us first set $q_1 = q_2 = 0$,
then the wave function in the domain (\ref{sfdomain}) behaves near the
horizon as
\begin{align}
 U(z) &\propto \frac{1}{ \sqrt{z}}  
  \cos \left(2 \sqrt{- \lambda \mu}z - \frac12 \pi (\nu + \frac12) \right)
  ~, & e^{i (q_0 + q_3)} &= 
  - \frac{\sin \pi  \left(\frac{k_0}{2} - \frac{\nu}{2}\right)}
         {\sin  \pi \left(\frac{k_0^*}{2} - \frac{\nu}{2}\right)} ~,
\end{align}
for $z>0$, and
\begin{align}
 U(z) &\propto \frac{1}{ \sqrt{z}}  
  \cos \left(2 \sqrt{- \lambda \mu}z - \frac12 \pi (\nu' + \frac12) \right)
  ~, & e^{i (q_0 - q_3 + 2 {\rm Im}\, k_0)} &= 
  - \frac{\sin \pi \left(\frac{k_0}{2} - \frac{\nu'}{2}\right)}
         {\sin \pi \left(\frac{k_0^*}{2} - \frac{\nu'}{2}\right)} ~,
\end{align}
for $z<0$. In this case, there is no interaction between the patches
with $z>0$ and $z<0$, and the parameters $q_0 \pm q_3$ fix the phase
factors of the boundary behaviors for the each patch.
In other words, there are no contributions to the each term of
the boundary condition (\ref{BC}), and hence $q_1,q_3$ parametrize
how the cancelation occurs inside the each patch $z>0$ or $z<0$.

For $q_i = 0 \, (i \neq 1)$  the asymptotic behavior is fixed by a
linear combination of
\begin{equation}
 \begin{split}
 U (z) &\propto 
 \begin{cases}
   J_{k_0} (2 \sqrt{- \lambda \mu}z) 
    + e^{i q_1} J_{k_0^*} (2 \sqrt{- \lambda \mu}z) &{\rm for} ~ z>0 ~,\\
   e^{-k_0 \pi i} J_{k_0} (2 \sqrt{- \lambda \mu}z) 
 + e^{i(q_1 - k_0^* \pi)} J_{k_0^*} (2 \sqrt{- \lambda \mu}z ) 
    &{\rm for} ~ z<0 ~,
 \end{cases}\\
 U (z) &\propto
 \begin{cases}
   J_{k_0} (2 \sqrt{- \lambda \mu}z) 
  + e^{-i q_1} J_{k_0^*} (2 \sqrt{- \lambda \mu}z) &{\rm for} ~ z>0 ~,\\
   -  \left(e^{-k_0 \pi i}J_{k_0}(2 \sqrt{- \lambda \mu}z) 
   + e^{- i(q_1 + k_0^* \pi)} J_{k_0^*}(2 \sqrt{- \lambda \mu}z) \right)
   &{\rm for} ~ z<0 ~,
 \end{cases}
\end{split}
\end{equation}
and for $q_i = 0 \, (i \neq 2)$
\begin{equation}
 \begin{split}
 U (z) &\propto 
 \begin{cases}
   J_{k_0}(2 \sqrt{- \lambda \mu}z)  
   + e^{i q_2} J_{k_0^*}(2 \sqrt{- \lambda \mu}z)  &{\rm for} ~ z>0 ~,\\
   i\left(e^{-k_0 \pi i}J_{k_0} (2 \sqrt{- \lambda \mu}z) 
   + e^{i(q_2 - k_0^* \pi)} J_{k_0^*}(2 \sqrt{- \lambda \mu}z) \right) 
  &{\rm for} ~ z<0 ~,
 \end{cases} \\
 U (z) &\propto
 \begin{cases}
   J_{k_0} (2 \sqrt{- \lambda \mu}z) 
   + e^{-i q_2} J_{k_0^*} (2 \sqrt{- \lambda \mu}z) &{\rm for} ~ z>0 ~,\\
   - i\left(e^{-k_0 \pi i}J_{k_0} (2 \sqrt{- \lambda \mu}z) 
   + e^{- i(q_2 + k_0^* \pi)} J_{k_0^*}(2 \sqrt{- \lambda \mu}z) \right) 
  &{\rm for} ~ z<0 ~.
 \end{cases}
\end{split}
\end{equation}
In these two cases, there are interactions between two patches, and the
corresponding parameters represent how the currents flow across the
horizons, or how cancelation occurs between the contributions from 
$z=\pm \infty$ parts in the boundary condition (\ref{BC}).

The different self-adjoint extensions correspond to different physics, and
we shall adopt the self-adjoint extensions suitable to our purpose.
Now that we have the eigenfunctions of the Hamiltonian $H_{ws}$ such as
(\ref{DJ}), (\ref{CJ}), (\ref{DJ2}) and (\ref{CJ2}), we require that
the domain (\ref{sfdomain}) includes the eigenfunctions as many as
possible. Later we check its physical relevance.
Wave functions in the domain of self-adjoint Hamiltonian must
satisfy the boundary condition (\ref{BC}), and we can see which
eigenfunctions are included in the domain by examining the boundary condition
with $U_1(z)$ as the eigenfunctions concerned and $U_2(z)$ as
(\ref{theta}). This is enough because the boundary values come from
only (\ref{theta}) for wave functions in the domain (\ref{sfdomain}) as
mentioned above. 
However, we will examine in this way later, and first we check by using
the wave functions (\ref{DJ}) and (\ref{DJ2}) as $U_1(z)$ and $U_2(z)$
in order to see how we can obtain a maximum set of eigenfunctions in the
domain. 
After that we will move to the case with the wave functions (\ref{CJ}) and
(\ref{CJ2}).

For the wave functions (\ref{DJ}) and (\ref{DJ2}) we use the following
notation 
\begin{equation}
 U_{2j-1} (z)
 =  \begin{cases} 
    J_{2j-1} (2 \sqrt{- \lambda \mu} z) & {\rm for}~ z>0 ~,\\
    f(j) e^{- (2j-1) \pi i}J_{2j-1} (2 \sqrt{- \lambda \mu} z)
    &{\rm for} ~ z<0 ~,
    \end{cases}
\label{ansatz}
\end{equation} 
where $f(j) \in \bc$ is a function of $j$.
The total normalization can be taken arbitrary and only
the relative factor is important.
Utilizing the analytic continuation of the Bessel function
\begin{equation}
 J_{\nu}(e^{\pi i}z) = e^{\nu \pi i }  J_{\nu}(z) ~,
\end{equation}
the boundary condition (\ref{BC}) reduces to
\begin{equation}
 \left(1+f(j_2)^* f(j_1) \right) \sin(\pi (j_1 - j_2))=0 ~.
\end{equation}
The above equation implies that if $j_1 -j_2 \in \bz$, or equivalently
$j_i=m_i + j_0 + 1/2$ with $(m_i = 0,1,2,\cdots, 0 < j_0 \leq 1)$,  
then the boundary condition (\ref{BC}) is satisfied irrespective of
$f(j)$. In other words, we can choose the normalization
of (\ref{DJ2}) independent to (\ref{DJ}).

In fact, the domain can include more eigenfunctions by carefully
choosing $f(j)$. Here we assign the forms of $f(j)$ as
\begin{align}
f(j) = \exp((\delta_c  + (2j-1) )\pi i) ~,
\label{ansatz2}
\end{align}
with $0 \leq \delta_c < 2 \pi$, then we have $j_1 - j_2 \in (\bz + 1/2)$ 
as well as $j_1 -j_2 \in \bz$, which leads to $j_i = 1/2(m_i + \nu_0 + 1)$ 
$(m_i = 0,1,2\cdots, 0 < \nu_0 \leq 1)$.
If we use $U_1=U_{2j-1}$ with (\ref{ansatz2}) and $U_2= U_M$
(\ref{theta}), then the boundary condition (\ref{BC}) reduces to the
following two equations
\begin{align}
 &(d_1^* + e^{(\delta_c + \nu_0) \pi i} e_1^*) 
   \sin \pi \left(\frac{\nu_0}{2} - \frac{k^*_0}{2} \right) +
  (d_2^* + e^{(\delta_c + \nu_0) \pi i} e^*_2) 
   \sin \pi \left(\frac{\nu_0}{2} - \frac{k_0}{2} \right) = 0 ~, \nonumber\\
 &(d_1^* - e^{(\delta_c + \nu_0) \pi i} e_1^*) 
   \cos \pi \left(\frac{\nu_0}{2} - \frac{k^*_0}{2} \right) +
  (d_2^* - e^{(\delta_c + \nu_0) \pi i} e^*_2) 
   \cos \pi \left(\frac{\nu_0}{2} - \frac{k_0}{2} \right) = 0 ~.
\end{align}
The solution to the above equations are written by using an unitary
matrix as 
\begin{align}
  & \qquad \begin{pmatrix}
    d_2^* \\ e_2^*
  \end{pmatrix}
   = \begin{pmatrix}
    M_{11} & M_{12} \\ M_{21} & M_{22}
  \end{pmatrix}
 \begin{pmatrix}
    d_1^* \\ e_1^*
  \end{pmatrix} ~, \\
 &M_{11} = M_{22} = - \frac12 
  \left( \frac{\sin \pi \left(\frac{\nu_0}{2} - \frac{k^*_0}{2} \right)}
     {\sin \pi \left(\frac{\nu_0}{2} - \frac{k_0}{2} \right)} +
         \frac{\cos \pi \left(\frac{\nu_0}{2} - \frac{k^*_0}{2} \right)}
     {\cos \pi \left(\frac{\nu_0}{2} - \frac{k_0}{2} \right)}\right) ~, \\
 &e^{- (\delta_c + \nu_0) \pi i} M_{12} 
  = e^{(\delta_c + \nu_0) \pi i} M_{21} = - \frac12 
  \left( \frac{\sin \pi \left(\frac{\nu_0}{2} - \frac{k^*_0}{2} \right)}
     {\sin \pi \left(\frac{\nu_0}{2} - \frac{k_0}{2} \right)} -
         \frac{\cos \pi \left(\frac{\nu_0}{2} - \frac{k^*_0}{2} \right)}
     {\cos \pi \left(\frac{\nu_0}{2} - \frac{k_0}{2} \right)}\right) ~.
  \nonumber
\end{align}
Thus, we can see that the self-adjoint parameters $q_i \, (i=0,1,2,3)$
are fixed by two parameters $\delta_c,\nu_0$.
From now on, we set $\delta_c = 0$ since the constant phase factor is
not relevant to the following discussions.
We do not choose a particular $\nu_0$ because
it relates to the label $j=1/2(m + \nu_0 + 1)$.  

In summary, the domain
we adopt has one parameter $\nu_0$, and the domain $D_{\nu_0}(H_{ws})$
for {\sl each} choice of $\nu_0$ comprises wave functions taking
discrete value $j=1/2(m+\nu_0+1)$ as\footnote{This result is essentially
the same as that in \cite{reduced2}.}
\begin{equation}
 \Phi_{m + \nu_0}^{\lambda , \mu}
 =  \sqrt{ 2 \pi (n + \nu_0)}  e^{i\lambda\gamma + i\mu\bar\gamma}
   z \, J_{m+\nu_0} \left(2 \sqrt{- \lambda \mu} z\right) ~,
\label{WDJ}
\end{equation}
for the both $z>0$ and $z<0$. The normalization is chosen 
so that the two point function becomes
\begin{equation}
 \langle \Phi_{m + \nu_0}^{\lambda_2 , \mu_2} ,
         \Phi_{n + \nu_0}^{\lambda_1 , \mu_1} \rangle
 = (2\pi)^3 \delta(\lambda_1 - \lambda_2) \delta(\mu_1 - \mu_2)
  \delta_{m,n} ~.
\end{equation}
These wave functions describe `bound states' in the attractive
Liouville potential. 

Let us move to the wave functions of the type (\ref{CJ}) and (\ref{CJ2})
\begin{equation}
 U_{2 i \omega} (z)
 =  \begin{cases} 
    \alpha_{\omega} J_{-2i\omega} (2 \sqrt{- \lambda \mu} z) +
    \beta_{\omega} J_{2i\omega} (2 \sqrt{- \lambda \mu} z)  
    &{\rm for} ~ z>0 ~,\\
    \gamma_{\omega} J_{-2i\omega} (2 \sqrt{- \lambda \mu} z) +
    \delta_{\omega} J_{2i\omega} (2 \sqrt{- \lambda \mu} z) 
    &{\rm for} ~ z<0 ~,
    \end{cases}
\label{CJ3}
\end{equation}
where we have changed the labels $a_3,a_4,b_3,b_4$ into 
$\alpha_{\omega},\beta_{\omega},\gamma_{\omega},\sigma_{\omega}$
expressing $j=1/2 + i \omega$ dependence explicitly.
As before, we check the boundary condition (\ref{BC})
with $U_1 = U_{2 i \omega}$ and $U_2 = U_M$ (\ref{theta}). 
The condition restricts the coefficients 
$\alpha_{\omega},\beta_{\omega},\gamma_{\omega},\sigma_{\omega}$ to%
\footnote{It is easier to examine the boundary condition (\ref{BC}) with
$U_1 = U_{2 i \omega}$ and $U_2=U_{2j-1}$ (\ref{ansatz})
obtained above to get the restriction of the coefficients. 
The results are the same.}
\begin{align}
&\alpha_{\omega} = a_{\omega} e^{-\pi (\omega + i \nu_0/2)} 
                - b_{\omega} e^{\pi (\omega + i \nu_0/2)}  ~,\nn
&\beta_{\omega} =  b_{\omega} e^{- \pi (\omega - i \nu_0/2)}
                - a_{\omega} e^{\pi (\omega - i \nu_0/2)}  ~,\nn
&\gamma_{\omega} = e^{- \pi (2 \omega - i \nu_0)}
                \left( b_{\omega} e^{-\pi (\omega + i \nu_0/2)} 
                    - a_{\omega}e^{\pi (\omega + i \nu_0/2)} \right)~,\nn
&\delta_{\omega}= e^{ \pi (2 \omega + i \nu_0)}
                \left( a_{\omega} e^{-\pi (\omega - i \nu_0/2)} 
                    - b_{\omega}e^{\pi (\omega - i \nu_0/2)} \right)  ~,
\label{ab}
\end{align}
with two arbitrary functions $a_{\omega}, b_{\omega}$.
This result implies that we can express $U_{2i\omega}$ by a linear
combination of two independent solutions of the form (\ref{ab}).
We will use the following two types of basic solutions; (1)
$(a_{\omega},b_{\omega}) = (1,0)$ and $(0,1)$, which are suitable to see
physical meaning, and (2) $(a_{\omega},b_{\omega}) = (1,1)$ and
$(1,-1)$, which are suitable to see mathematical meaning.

(1) $(a_{\omega},b_{\omega}) = (1,0)$ and $(0,1)$.
Let us first see the wave function with $(a_{\omega},b_{\omega}) = (1,0)$.
Notice that the coefficients may be obtained by
analytic continuation up to total normalization as for $z>0$
\begin{equation}
 K_{2 i \omega}\left(e^{- \pi i/2} 2 \sqrt{-\lambda \mu} z\right) =
 \frac{\pi}{2}\frac{e^{- \pi\omega} J_{-2i\omega}
 ( 2 \sqrt{-\lambda \mu} z )
 - e^{\pi\omega} J_{2i\omega} ( 2 \sqrt{-\lambda \mu} z ) 
   }{\sin(2 i \pi \omega)}~,
\label{analytic}
\end{equation}
and for $z < 0$ 
\begin{equation}
 K_{2 i \omega}\left(e^{ \pi i/2} 2 \sqrt{-\lambda \mu} z\right) =
 \frac{\pi}{2}\frac{e^{ \pi\omega} J_{-2i\omega}( 2 \sqrt{-\lambda \mu} z)
 - e^{- \pi\omega} J_{2i\omega}( 2 \sqrt{-\lambda \mu} z) 
   }{\sin(2 i \pi \omega)}~.
\end{equation}
For $\lambda \mu >0$ case the wave function with (\ref{K}) belongs to the
Hilbert space since it behaves well near the horizon $z \to \infty$ as
\begin{equation}
 \Psi_{2i\omega} \left(2 \sqrt{\lambda \mu} z\right) \sim \sqrt{z}
  \exp (- i E t + i L \theta - \sqrt{L^2 - E^2}z) ~,
\end{equation}
where $-\lambda + \mu = E$ and $\lambda + \mu = L$ represent energy and
angular momentum, respectively. That is to say, there is a damping
factor $\sim e^{- \sqrt{L^2 - E^2} z}$ in the wave functions. 
Using the analytic continuation, we have 
\begin{equation}
 \Psi_{2i\omega} \left(e^{-\pi i/2} 2 \sqrt{- \lambda \mu} z\right) 
 \sim \sqrt{z}
  \exp (- i E t + i L \theta + i\sqrt{E^2 - L^2}z ) ~.
\end{equation}
In general, the wave function may be represented as a linear combination
of in-coming and out-going plane waves, but the analytic continuation
lead to only an out-going plane wave 
$\sim e^{-i E t + i\sqrt{L^2 - E^2} z}$ near the horizon $z \to \infty$.
Similarly, in the $z<0$ patch, the analytic continuation leads to only
an in-coming plane wave, so the wave function with
$(a_{\omega},b_{\omega}) = (1,0)$ shows the current flow from the 
the $z>0$ patch to the $z<0$ patch.
This current flow is conserved because the wave functions satisfy the
boundary condition (\ref{BC}) (even though the boundary condition
originates from the self-adjointness condition of the minisuperspace
Hamiltonian $H_{ws}$ (\ref{H_ws})).
For the case with $(a_{\omega},b_{\omega})=(0,1)$, the $z<0$ patch has an
out-going plane wave and the $z>0$ patch has an in-coming plane wave. 
This wave function implies the conserved current flow from the $z<0$
patch to the $z>0$ patch. 
Since the wave functions in the domain can be written as a linear
combination of the two solutions, we can follow the conserved current
flow through the successive patches. 

In the minisuperspace model viewpoint, wave functions mean the
probability of existence. Also in this context, we can show in the
similar way that the wave functions in this basis imply the conserved
probability flow with respect to the worldline time $\tau$
(see \cite{reduced2} for a discussion on the probability flow).

(2)  $(a_{\omega},b_{\omega}) = (1,1)$ and $(1,-1)$. We first use the
normalization of the wave functions as
\begin{equation}
\Phi_{ \nu_0,\omega}^{(n) \lambda , \mu} =  
 \sqrt{\frac{2 \pi \omega}{\sinh 2 \pi \omega}}
 e^{i\lambda\gamma + i\mu\bar\gamma} z \,
 U^{(n)}_{\nu_0,\omega} (z) ~.
\label{WCJ}
\end{equation}
We denote the wave functions with  
$(a_{\omega},b_{\omega}) = (1,1)$ and $(1,-1)$ as 
$\Phi_{ \nu_0,\omega}^{(0)\lambda , \mu}$ and 
$\Phi_{ \nu_0,\omega}^{(1)\lambda , \mu}$, respectively. 
We have also defined
\begin{equation}
U^{(n)}_{\nu_0,\omega} = 
 \begin{cases} 
 J_{- 2 i \omega}(2\sqrt{-\lambda\mu}z) + \Theta^{(n)}_{\nu_0} (\omega)
      J_{2 i \omega}(2\sqrt{-\lambda\mu}z) & {\rm for } ~ z>0 ~,\\
 e^{\pi i (\nu_0 + n)}\left(
  J_{- 2 i \omega}(2\sqrt{-\lambda\mu}e^{-\pi i}z) 
  + \Theta^{(n)}_{\nu_0} (\omega)
      J_{2 i \omega}(2\sqrt{-\lambda\mu} e^{-\pi i}z)\right) 
   & {\rm for } ~ z<0 ~,
 \end{cases}
\end{equation}
where 
\begin{equation}
  \Theta^{(n)}_{\nu_0} (\omega) = \frac{\sinh \pi (\omega - i (\nu_0 + n)/2 )}
              {\sinh \pi (\omega + i (\nu_0 + n)/2)} ~.
\end{equation}
In this normalization, we can check by closely following the appendix
of \cite{mini2} that the wave functions in the
domain  $D_{\nu_0}(H_{ws})$  satisfy the completeness condition such as
\begin{align}
 &\sum_{m = 0}^{\infty}
  \left(\Phi^{\lambda,\mu}_{m+\nu_0} (z,\gamma,\bar\gamma)\right)^* 
   \Phi^{\lambda,\mu}_{m+\nu_0} (z',\gamma,\bar\gamma)  \nonumber \\
 &\qquad \qquad +  \sum_{n=0,1} \int_0^{\infty} d \omega
   \left(\Phi^{(n)\lambda,\mu}_{\nu_0,\omega}   (z,\gamma,\bar\gamma)\right)^*
     \Phi^{(n)\lambda,\mu}_{\nu_0,\omega}(z',\gamma,\bar\gamma)
    = 2 \pi |z|^3 \delta(z - z') ~,
\label{complete}
\end{align}
as well as the orthogonality condition. This means that an arbitrary
function defined in the patches with $z>0$ and $z<0$ can be decomposed
by the set of the wave functions in the domain $D_{\nu_0}(H_{ws})$ for
each label $\nu_0$.
If we restrict ourselves to the patch with $z>0$, then the each
set of wave functions
$(\Psi^{\lambda,\mu}_{2n+\nu_0},\Psi^{(0)\lambda,\mu}_{\nu_0,\omega})$ 
and
$(\Psi^{\lambda,\mu}_{2n+1+\nu_0},\Psi^{(1)\lambda,\mu}_{\nu_0,\omega})$ 
with $n = 0,1,2,\cdots$ and $\omega > 0$ satisfies the orthogonal and
complete conditions by itself.
There are two sets of orthogonal and complete bases because we have
to describe the functions defined in the both patches with $z>0$ and
$z<0$.

In spite of this fact, we will use a different normalization as
\begin{equation}
\Phi_{ \nu_0,\omega}^{(n) \lambda , \mu} = 
 \frac{(- \lambda \mu)^{i \omega}}{\sqrt2} \Gamma(1-2i \omega)
 e^{i\lambda\gamma + i\mu\bar\gamma} z \, U^{(n)}_{\nu_0, \omega} (z) 
\label{WCJ2}
\end{equation}
for later convenience. Here again, we chose the normalization so
that the two point function and asymptotic behavior of the wave
function take the forms of (\ref{2pt}) and (\ref{AB}),
respectively. The corresponding reflection amplitude is readily
computed:
\begin{equation}
R(\omega) =  (- \lambda \mu)^{2i\omega}
 \frac{\Gamma(1-2i \omega)}{\Gamma(1+2i\omega)} \,
  \Theta^{(n)}_{\nu_0} (\omega)~.
\label{RA2}
\end{equation}

Before ending this section, let us comment on the validity
 of the choice of the domain $D_{\nu_0}(H_{ws})$, namely the
choice of the self-adjointness parameters. 
First, the set of eigenfunctions in the domain
gives orthogonal and complete basis as mentioned above. 
Second, the wave functions in the domain $D_{\nu_0}(H_{ws})$ with
$\nu_0=1$ reproduces the wave functions in the single cover of $AdS_3$
(see, for example, \cite{ST1}).\footnote{In the single cover of $AdS_3$
 the wave functions must be the same in the all $z<0$ patches by
 definition. If we require that the results with $\arg z = \pi$, which was
 obtained in the context, and those with $\arg z = -\pi$, which can be
 analyzed in a similar way, are the same, then we find the requirement
 reduces to $e^{2 \pi i \nu_0} = 1$ or $\nu_0 = 1$.} 
In particular, for $j>1/2$ states, the label $j$ takes a half-integer
 value as in the single cover case.
Finally, there is a probability flow between the Poincar\'e patches in the
choice, and this is consistent with the fact that in the global
coordinates the probability flows from infinite past to infinite
future.


\section{Primary fields and correlation functions}
\label{coordinate}

In the previous section, we constructed the domain $D_{\nu_0}(H_{ws})$
of the selfadjoint Hamiltonian. 
The eigenstates in the domain correspond to the normalizable states in
$SL(2.\br)$ WZNW model. 
Basic quantities in WZNW model are correlation functions of primary
fields, and fields correspond to normalizable states as 
\begin{equation}
 \ket{\lambda,\mu,j} = \lim_{w \to 0}
 \Psi^{\lambda,\mu}_{j} (w,\bar w) \ket{0} ~.
\label{state-op}
\end{equation} 
The ket $ \ket{\lambda,\mu,j}$ at the left hand side represents a
normalizable state, and the field $ \Psi^{\lambda,\mu}_{j}$
corresponds to the state when acting to a vacuum $\ket{0}$.
The worldsheet is described by the coordinates $w=e^{\tau+\sigma}$ and
$\bar w = e^{\tau-\sigma}$, and in the minisuperspace limit, we neglect
the $\sigma$-dependence.

Moreover, in the quantum mechanics, the
operator corresponding to the primary field can be found from the
properties under the $SL(2,\br)$ transformation as
\begin{equation}
 \Psi^{\lambda,\mu}_{j} \Phi^{\lambda',\mu'}_{j'}(z,\gamma,\bar \gamma) 
  := \Phi^{\lambda,\mu}_{j}(z,\gamma,\bar \gamma) 
      \Phi^{\lambda',\mu'}_{j'}(z,\gamma,\bar \gamma) ~.
\end{equation}
Therefore, the minisuperspace analogy of the multi-point correlation
functions are given by the overlaps of wave functions
\begin{equation}
 \left\langle \prod_{i=1}^N \Phi^{\lambda_i,\mu_i}_{j_i}
 (z,\gamma,\bar \gamma) \right\rangle
 := \int d\gamma d \bar \gamma d z |z|^{-3}  \, \prod_{i=1}^N
 \Phi^{\lambda_i,\mu_i}_{j_i} (z,\gamma,\bar \gamma) ~.
\end{equation}

In this section, we construct primary fields and compute
correlation functions by making use of the $SL(2,\br)$ symmetry.  
In order to do so, it is easier to use a different basis from
the previous one. It is given by a
Fourier transform of the previous one
\begin{eqnarray}
\Psi^j_A(x,\bar x;g)= \frac{1}{2\pi^2} \int d \lambda d \mu \, e^{- i
\lambda x - i \mu \bar x}  \Phi^{\lambda, \, \mu}_j(g)  ~.
\label{inverse}
\end{eqnarray}
This basis is convenient because the action of $SL(2,\br)$ Lie algebra 
are generated by differential operators, and we can see how much the
$SL(2,\br)$ symmetry fixes functional form of the primary fields or
correlation functions.

Since we construct the primary fields with well behaviors under the
$SL(2,\br)$ action, the primary fields satisfy differential
equations. In the Euclidean theory ($SL(2,\bc)/SU(2)$ WZNW model), 
the solution to the differential equations is unique up to normalization 
(see, e.g., \cite{Teschner1}). However, in our Lorentzian theory, the
solution to the differential equations is not unique, and given in a
linear combination of two independent solutions.
The different solution corresponds to the different wave function as in 
(\ref{CJ}) with different $a_3,a_4$, and a fixed
wave function corresponds to a particular linear combination of two
solutions due to the state-operator correspondence.  
The situation is quite similar also for two point functions
and three point functions. 
These correlation functions obey differential equations because of
the $SL(2,\br)$ invariance,
and in the Euclidean theory, the solutions to the differential equations
are almost unique, and the task is only to fix one coefficient for each
\cite{Teschner1}. However, in the Lorentzian theory, the solutions to
the differential equations are less restricted, and we have to compute more
coefficients. In the minisuperspace limit, the correlation functions can
be computed as overlaps of wave functions, so we can fix these
coefficients. For the full CFT case, we may have to use the $SL(2,\br)$
symmetry to compute the correlation functions.

As seen in the previous section, it is essential to use the both patches
with the coordinate $z>0$ and $z<0$ for constructing the self-adjoint
extension. However, in this section we only consider a patch with $z>0$,
and for this reason we will mainly use the coordinate $\phi = - \log z $.
It is enough for the purpose to see how much the $SL(2,\br)$ symmetry
fixes the forms of primary fields and correlation functions. 
Of course, it is necessary to consider also the patch
with $z<0$ in order to construct primary fields defined in the whole
spacetime and to compute full correlation functions. It can be done by
using the similar analysis, and the results may include some additional
factors.

\subsection{Primary fields and $SL(2,\br)$ symmetry}

Consider again the $SL(2, \br)$ WZNW action.
In the minisuperspace limit, the action is reduced to
\begin{equation}
 S = \frac{k}{8} \int d \tau {\rm Tr} (g^{-1} \partial_{\tau} g)^2 ~.
\end{equation}
In the expression, invariance under the transformation
$g \to h_L g h_R$ with $h_{L,R} \in SL(2,\br)_{L, R}$ is manifest.
In the parametrization (\ref{para}), the currents associated with the
symmetry are given by
\begin{align}
 J^-_L &= \partial_{\gamma}~,&
 J^3_L &= \gamma \partial_{\gamma} - \frac{1}{2} \partial_{\phi}~,&
 J^+_L &= \gamma^2 \partial_{\gamma} - \gamma \partial_{\phi}
         - e^{-2 \phi} \partial_{\bar \gamma}~, \nonumber \\
 J^-_R &= \partial_{\bar\gamma}~,&
 J^3_R &= \bar\gamma \partial_{\bar\gamma} - \frac{1}{2} \partial_{\phi}~,&
 J^+_R &= \bar\gamma^2 \partial_{\bar\gamma} - \bar\gamma \partial_{\phi}
         - e^{-2 \phi} \partial_{\gamma} ~.
\end{align}
As mentioned above, we shall use the parametrization
$(x, \bar x)$ instead of $(\lambda,\mu)$ (in addition to $j$).
The variables $x$ and $\bar x$ are real-valued, so they are not related
each other by complex conjugation. 
In the parametrization, the action of
$SL(2,\br)_L \times SL(2,\br)_R$ Lie algebra can be generated by the
following set of differential operators as
\begin{align}
 D^-_L &= -\partial_x~,&
 D^3_L &= -x \partial_x - (j+1) ~,&
 D^+_L &= - x^2 \partial_x - 2(j+1) x ~, \nonumber \\
 D^-_R &= -\partial_{\bar x}~,&
 D^3_R &= -\bar x \partial_{\bar x} - (j+1) ~,&
 D^+_R &= - {\bar x}^2 \partial_{\bar x} - 2(j+1) {\bar x} ~,
\label{diffop}
\end{align}
where, as in (\ref{KG}), $c_2=j(j-1)$ refers to the Casimir invariance
of the $SL(2, \br)$ Lie algebra.

In the $SL(2,\br)$ WZNW model, primary fields are labeled by the
representation of the $SL(2,\br)$ Lie algebra. 
In the $(x,\bar x)$ basis, the primary
fields with the label $j$ transform as
\begin{eqnarray}
J_{L,R}^a \Psi^j(x,\bar x ; g)= D_{L,R}^a  \Psi^j(x,\bar x;g)
\qquad \quad (a=\pm,3) ~.
\label{diffeq}
\end{eqnarray}
The solution to these equations is given locally by
\begin{equation}
\Psi^j(x, \bar x; g) \sim [F(x,\bar x;g)]^{-2j}
\qquad \mbox{where} \qquad
F(x,\bar x;g) \equiv (\gamma - x)(\bar\gamma-\bar x)e^{\phi}+ e^{-\phi} ~.
\label{Fxx}
\end{equation}
However, the solution has a singularity at $F(x,\bar x;g) = 0$, 
and needs to be prescribed suitably at such loci.

What about the Euclidean $AdS_3$? In this case, the dynamics is
described by the $SL(2,\bc)/SU(2)$ WZNW model, and the corresponding
wave function $F(x,\bar x;g)$ is manifestly positive-definite, and hence
(\ref{Fxx}) is the unique solution up to normalization. 
For the Lorentzian $AdS_3$, however, there are singularities at
$F(x,\bar x;g)= 0$, yielding
\begin{equation}
 \Psi_{A}^j(x,\bar x ; g) =
 \begin{cases}
  \, C_j \, |F(x,\bar x;g)|^{-2j} & \qquad {\rm for} \quad ~F(x,\bar x;g) 
  >0 ~,\\
  A_j  C_j
  \, |F(x,\bar x;g)|^{-2j} & \qquad {\rm for} \quad ~F(x,\bar x;g) <0 ~.
   \end{cases}
\label{wfxx}
\end{equation}
Here $C_j$ is an overall normalization factor, while $A_j$ is a
relative phase factor.

To appreciate how $A_j$ is determinable, consider as an example
the analytic continuations, viz. $\pm i \varepsilon$ prescription. 
One readily finds that the primary field is given by:
\begin{equation}
 \Big[
 \frac{1}{F(x, \bar x; g) \pm i \varepsilon }\Big]^{2j}
 =  \begin{cases}
  |F(x,\bar x;g)|^{-2j} & \quad {\rm for} ~F (x,\bar x;g) >0 ~,\\
  e^{\mp 2 \pi i j }
  | F(x,\bar x;g)|^{-2j} & \quad {\rm for} ~F (x,\bar x;g) <0 ~,
   \end{cases}
\label{wick}
\end{equation}
hence the phase-factor is determined as $A_j= \exp (\mp 2 \pi i j)$.
It can be said that the general primary field (\ref{wfxx}) is given by a
linear combination of the basic primary fields given by the two different
analytic continuations.

As an another example, consider wave functions with definite parity 
($(-1)^{2 \epsilon}$ with $\epsilon = 0,1/2$ mod $1$).
To extract the parity transformation rules, we find it convenient
to express the function $F$ as
\begin{eqnarray}
   F(x,\bar x;g) =
   \begin{pmatrix}
    1 & -x
   \end{pmatrix}
     g
   \begin{pmatrix}
    1 \\ -\bar x
   \end{pmatrix} ~.
\end{eqnarray}
Then, the action of $T_h: g \to h_L g h_R$ on the primary field of
definite parity is given by
\begin{align}
 &T_h \Psi_{(-1)^{2 \epsilon}}^j(x,\bar x;g)
 \equiv \Psi^j_{(-1)^{2 \epsilon}}(x,\bar x; h_L^{-1} g h_R^{-1})
  \nonumber \\
  & = | (\beta x + \delta) (\bar \beta x + \bar \delta) |^{-2j}
    [{\rm sgn}
   ((\beta x + \delta) (\bar \beta x + \bar \delta))]^{2 \epsilon}\,
      \Psi_{(-1)^{2 \epsilon}}^j
  \left(\frac{\alpha x + \gamma}{\beta x + \delta},
  \frac{\bar \alpha \bar x + \bar \gamma}{\bar \beta \bar x + \bar \delta};
  g \right), \label{parity}  \end{align}
where
\begin{align}
 h_L &= \begin{pmatrix}
    \alpha & \lambda \\ \beta & \delta \end{pmatrix} ~,&
  h_R &= \begin{pmatrix} \bar \alpha & \bar \lambda \\
      \bar \beta & \bar \delta \end{pmatrix} ~. 
\end{align}
Therefore, from the parity transformation $g \to - g$, 
we can see that the wave function with parity $+1$ is given by 
\begin{equation}
 \Psi_{+1}^j(x, \bar x; g) \sim  |F(x,\bar x;g)|^{-2j} ~,
\label{waveftn}
\end{equation}
or (\ref{wfxx}) with $A_j=+1$ and the one with parity $-1$ is given by
\begin{eqnarray}
\Psi_{-1}^j(x, \bar x; g) \sim  {\rm sgn} (F(x,\bar x;g)) 
   |F(x,\bar x;g)|^{-2j} ~. 
\label{waveftn2}
\end{eqnarray}
or (\ref{wfxx}) with $A_j=-1$.
The general primary field with $A_j$ can be written as a linear
combination of the primary fields with parity $\pm 1$, and this fact
will be found useful to compute the three point functions.

\subsection{Relation to the normalizable states}
\label{2ptxx}

Because of the state-operator correspondence, we can compare the
primary fields defined above with the eigenfunctions included in the domain
$D_{\nu_0}(H_{ws})$. To see the relation, we have to take a care about
the difference of the bases. In fact, the primary fields in the 
$(x,\bar x)$ basis (\ref{wfxx}) are Fourier transforms of the wave
functions in the $(\lambda,\mu)$  basis\footnote{ We define the
integration measure as $ d x d \bar x \equiv d y d z$ with 
$x=y+z,\bar x = y-z$.}
\begin{eqnarray}
\Psi^{\lambda, \, \mu}_j(g) = \int d x d \bar x \, e^{i
\lambda x + i \mu \bar x} \Psi^j_A(x,\bar x;g) ~. \label{ftback}
\end{eqnarray}
Demanding them to match with the wave functions obtained in
section \ref{momentum}, we will be able to extract the `phase-factor' $A_j$.
The inverse of the above Fourier transform  (\ref{inverse}) is given by
integrating over the whole ranges of $\lambda$ and $\mu$, so $A_j$ should
not depend on the sign of $\lambda \mu$. 

We should remark that the above integral converges only
if $1/2 < {\rm Re}\, j < 3/4$. For $j=1/2 + i \omega$ case, we should
first introduce a regulator $j= 1/2 + \epsilon + i \omega$, and then take the
limit of $\epsilon \to + 0$. For $j>1/2$ case, we can only compare the two
representations within a small range $1/2 < j < 3/4$ to fix $A_j$, and
for $j \geq 3/4$ we will use the analytic continuation on $j$ as $A_j$.

Let us first consider $\lambda \mu > 0$ case.
This is the case already present in the Euclidean theory. 
We find that the integral (\ref{ftback}) becomes
\begin{align}
\Psi^{\lambda, \, \mu}_j(g) = C_j \left[ \frac{\cos \pi (1-2j)
    + A_j}{(2j -1) \sin \pi (1-2j)}\right]
    \frac{2 \pi (\lambda \mu)^{j-1/2}}{\Gamma (2j -1)}
      e^{i \lambda \gamma + i \mu \bar \gamma}
     e^{- \phi} K_{2 i \omega} \left(2 \sqrt{\lambda \mu} e
  ^{-\phi}\right) ~. \label{integral}
\end{align}
Hence, we recover precisely the result (\ref{WK}) {\sl provided}
we choose the normalization coefficient ($j=1/2 + i\omega$)
\begin{equation}
 C_j \equiv \frac{1}{\sqrt2 \pi}
  \left[ \frac{ \omega \sinh(2 \omega \pi)}{\cosh(2 \omega\pi)
   + A_j}\right]
  ~, \label{C_j}
\end{equation}
Here we should notice that there appears $A_j$ dependence only in the
total normalization.

On the other hand, in $\lambda \mu < 0$ case, there is no Euclidean
counterpart. 
The Fourier transform (\ref{ftback}) in the present case is
\begin{align}
  \Phi^{\lambda, \, \mu}_j (g)
 = C_j \frac{\pi^2 (- \lambda \mu)^{j-1/2}}{\Gamma (2j) \sin ^2 \pi (1 - 2j)}
 & e^{i\lambda\gamma + i\mu\bar\gamma} e^{-\phi}
 \left((\cos \pi (1 - 2j) + A_j) 
  J_{1-2j} \left(2\sqrt{-\lambda\mu}e^{-\phi}\right)
  \right. \nonumber\\ 
 & \left. - (\cos \pi (1- 2j) \cdot A_j +1)
 J_{2 j - 1}\left(2\sqrt{-\lambda\mu}e^{-\phi}\right)
 \right)  ~.
\label{Fourier>}
\end{align}
When $j=1/2 + i \omega$, the normalization is set by (\ref{C_j}) and we
have
\begin{align}
  \Phi^{\lambda, \, \mu}_\omega (g)
 &= \frac{(- \lambda \mu)^{i \omega}}{\sqrt2} \Gamma(1-2i \omega)
 e^{i\lambda\gamma + i\mu\bar\gamma} e^{-\phi} \nonumber \\
 & \qquad  \times
 \left(J_{- 2 i \omega}\left(2\sqrt{-\lambda\mu}e^{-\phi}\right)
  - \frac{\cosh(2\omega\pi)A_j+1}
    {\cosh(2\omega\pi)+A_j}
 J_{2 i \omega}\left(2\sqrt{-\lambda\mu}e^{-\phi}\right)
 \right)  ~.
\end{align}
Requiring that the result be (\ref{WCJ2}), we find the
following relation between  $\Theta^{(n)}_{\nu_0}$ and $A_j$:
\begin{align}
 \Theta^{(n)}_{\nu_0}(\omega) &=   - \frac{\cosh(2\omega\pi)A_j+1}
  {\cosh(2\omega\pi)+A_j}, &&{\rm equivalently},
 &A_j &=
 - \frac{\cosh \pi (\omega - i (\nu_0 + n)/2)}
        {\cosh \pi (\omega + i (\nu_0 + n)/2)} ~.
\label{A_omega}
\end{align}
As a checkpoint, take again the two examples considered in the
previous section. For these examples, the forms of 
$\Theta$ ($= \Theta^{(n)}_{\nu_0}$ in the above example) 
and $A_j$ are known already. For the example of analytic continuations, 
$\Theta = - \exp (\mp 2\omega\pi)$ (see (\ref{analytic})), and
(\ref{A_omega}) correctly reproduces $A_j = - \exp(\mp 2\omega\pi)$ in
(\ref{wick}). For the example of parity eigenstates,
$A_j=+1$ and $A_j=-1$. It then follows from (\ref{A_omega}) that
$\Theta =-1$ and $\Theta =+1$, respectively, and these are the wave
functions with definite parity as in \cite{SL(2R),ST1}.

For $1/2<j<3/4$ case, if we use in (\ref{Fourier>})
\begin{align}
 C_j &= \frac{\sqrt{2 \pi (2j-1)} \Gamma (2j)}
             {\pi^2 (-\lambda \mu)^{j-1/2}} ~, 
 &A_j &= - \cos \pi (1 - 2j) ~, 
\label{WDJp}
\end{align}
then we reproduce the wave function (\ref{WDJ}).
We can show as a consistency check that the integral (\ref{ftback}) for
$\lambda \mu > 0$ vanishes if we use (\ref{WDJp}).
For $j \geq 3/4$ case, we cannot perform the Fourier transform
(\ref{ftback}) because the integral diverges.
Therefore, we {\it define} the primary fields with $j\geq 3/4$
corresponding to the states by using the parameters (\ref{WDJp}) with
$j\geq 3/4$.

\subsection{Two point functions}

We mainly consider $j=1/2 + i\omega$ case.
For the primary fields  (\ref{wfxx}) with the coefficients (\ref{C_j})
and (\ref{A_omega}), asymptotic behavior near the boundary 
$z\rightarrow 0$, equivalently, $\phi \to \infty$ is extracted:
\begin{eqnarray}
 \Psi^j_A (x,\bar x;g) \sim \delta(\gamma -x)\delta(\bar \gamma -\bar x)
  e^{(2 j -2) \phi} + C_j
 |(\gamma -x)(\bar \gamma - \bar x)|_A^{-2j} e^{- 2j \phi} ~.
 \label{asympt}
\end{eqnarray}
Here, we abbreviated
\begin{equation}
 |(\gamma -x)(\bar \gamma - \bar x)|_A^{-2j} =
 \begin{cases}
 |(\gamma -x)(\bar \gamma - \bar x)|^{-2j} &
 {\rm for} ~  (\gamma -x)(\bar \gamma - \bar x)>0 ~,\\
  A_j |(\gamma -x)(\bar \gamma - \bar x)|^{-2j} &
 {\rm for} ~ (\gamma -x)(\bar \gamma - \bar x)<0 ~.
 \end{cases}
\end{equation}
The second term in (\ref{asympt}) represents the reflection
amplitude, and we have the reflection relation:
\begin{equation}
 \Psi^{j}_A(x,\bar x;g) = C_j  \int dx' d\bar x'
  |(x -x')(\bar x - \bar x')|_A^{-2j} \Psi^{1-j}_A (x',\bar x';g) ~.
  \label{reflectionrelation}
\end{equation}
This relation is verifiable by utilizing the Mellin
transformations. Because of this relation, we can restrict the spectrum
to $\omega > 0$, and this truncation is consistent with the completeness
condition (\ref{complete}).

The wave functions with $j>1/2$ represent the bound states, and hence
there is no such a reflection relation. This is consistent with the fact
that wave functions with $1-j$ are not included in the Hilbert
space.

Two point function can be extracted from the reflection relation
(\ref{reflectionrelation}) or from the Fourier transform of the
result in the previous section. Functional form of the two point
function obtained so is restricted by the 
$SL(2,\br)_L \times SL(2,\br)_R$ symmetry. 
Since the two point functions ought to be invariant under the
transformations generated by the differential operators (\ref{diffop}),
we obtain the following ansatz $(x_{12}\equiv x_1 - x_2)$:
\begin{align}
 &\left\langle \Psi^{j_2}_A(x_2,\bar x_2) \Psi^{j_1}_A (x_1,\bar x_1) 
 \right\rangle = N_{j_1}  \delta(j_1 + j_2 - 1)
  \delta(x_{12}) \delta(\bar x_{12}) \nonumber\\ & \qquad\qquad
    + \delta(j_1-j_2) \sum_{\eta,\bar \eta=0,1/2}
  D^{\eta \bar \eta}_{j_1} ({\rm sgn} \, x_{12})^{2\eta}
  ({\rm sgn} \, \bar x_{12})^{2\bar\eta}
  |x_{12}|^{-2j_1} |\bar x_{12}|^{-2j_1} ~.
\label{2pfform}
\end{align}
Notice that four possible solutions are emerging, as shown in the
second term, though there is only one solution possible in the
Euclidean theory.

Since the Fourier transform of the two point
function is already obtained in the previous subsection, the
coefficients $N_j$ and $D^{\eta,\bar\eta}_j$ can be determined as
\begin{align}
 &\left\langle \Psi_{A}^{\omega_2}(x_2,\bar x_2)
  \Psi^{\omega_1}_{A} (x_1,\bar x_1) \right\rangle 
   \nonumber \\ & \qquad \qquad
   = 4\pi \delta(x_{12}) \delta(\bar x_{12})
     \delta(\omega_1 + \omega_2)
  + 4\pi C_{\omega_1} |x_{12}\bar x_{12}|_A^{-2j}
 \delta(\omega_1-\omega_2) ~. 
\label{final}
\end{align}
It is quite significant that we find the {\sl identical} result
(including normalization factors) for both $\lambda \mu >0$ and
$\lambda \mu <0$ branches. Fourier transform of the second term in
(\ref{final}) yields precisely the reflection function $R(\omega)$
(\ref{RA}) and (\ref{RA2}) for $\lambda \mu
>0$ and $\lambda \mu <0$, respectively. 

For $j>1/2$ case, the two point function is simply given by
\begin{equation}
\left\langle \Psi^{n_2 + \nu_0}(x_2,\bar x_2)
  \Psi^{n_1 + \nu_0} (x_1,\bar x_1) \right\rangle
   = 4\pi \delta(x_{12}) \delta(\bar x_{12})
     \delta_{n_1,n_2}~.
\end{equation}
There is only one term involved, which is related to the fact
that there is no reflection relation in this case.

\subsection{Three point functions}
\label{3ptfn}

In this subsection, we compute three point functions.
As in the primary operators and the two point functions, the $SL(2,\br)$
symmetry restricts functional form of three point functions. 
However, if we only solve the differential equations coming from the
requirement of the $SL(2,\br)$ invariance, then the solution to the
equation has too many undetermined coefficients. Therefore, we also utilize
the properties under the parity transformation. 
In order to do this, we first decompose the primary field
(\ref{wfxx}) by the fields with definite parity $A=\pm1$
\begin{equation}
 \Psi^{j}_A (x,\bar x;g) =
   P^{0}_A \Psi_{+1}^j (x,\bar x;g)
 + P^{1/2}_A \Psi_{-1}^j (x,\bar x;g) ~.
\end{equation}
Then, the general three point function can be obtained as
\begin{equation}
  \left\langle  \prod_{i=1}^3 \Psi_{A_i}^{j_i}(x_i,\bar x_i)
 \right\rangle =
 \sum_{\begin{subarray}{c}
  \epsilon_1,\epsilon_2 = 0,1/2 \\
  \epsilon_3=\epsilon_1+\epsilon_2 \mod 1
 \end{subarray}}
 P^{\epsilon_1}_{A_1} P^{\epsilon_2}_{A_2} P^{\epsilon_3}_{A_3}
  \left\langle \prod_{i=1}^3
  \Psi_{(-1)^{2\epsilon_i}}^{j_i}(x_i,\bar x_i) \right\rangle
   ~.
 \label{lc3pt}
\end{equation}
Here we have used the parity conservation 
$\epsilon_1 + \epsilon_2 + \epsilon_3 = 0 \mod 1$.

The primary fields with definite parity transform under the $SL(2,\br)$
action as in (\ref{parity}). 
Therefore, by considering the three point functions
of primary fields of this type, we can deduce
functional form of the three point functions via these symmetries as
in \cite{SL(2R),ST1}:
\begin{equation}
 \left\langle  \prod_{i=1}^3 \Psi_{(-1)^{2\epsilon_i}}^{j_i}(x_i,\bar x_i)
 \right\rangle  = \sum_{\omega, \bar \omega = 0,1/2}
          C_{\omega,\bar \omega}(j_i)
          K_{\omega} (x_i) \bar K_{\bar \omega} (\bar x_i) ~,
\label{3pfform}
\end{equation}
where
\begin{equation}
\begin{split}
 &\qquad K_{\omega} (x_i) =
 ({\rm sgn}x_{12})^{2\eta_1}  |x_{12}|^{- \tilde j_1}
 ({\rm sgn}x_{23})^{2\eta_2}  |x_{23}|^{- \tilde j_2}
 ({\rm sgn}x_{31})^{2\eta_3}  |x_{31}|^{- \tilde j_3} ~,\\
 &\qquad \bar K_{\bar \omega} (\bar x_i)=
 ({\rm sgn}\bar x_{12})^{2\bar \eta_3}  |\bar x_{23}|^{- \tilde j_1}
 ({\rm sgn}\bar x_{23})^{2\bar \eta_2}  |\bar x_{23}|^{- \tilde j_2}
 ({\rm sgn}\bar x_{31})^{2\bar \eta_3}  |\bar x_{31}|^{- \tilde j_3} ~,
\end{split}
\end{equation}
\begin{equation}
 \eta_i = \epsilon_i + \omega \mod 1 ~, \quad
  \bar \eta_i = \epsilon_i + \bar \omega \mod 1~, \quad
 \tilde j = j_1 + j_2 + j_3 ~,\quad
  \tilde j_i =  \tilde j - 2 j_i ~.
\end{equation}
In the Euclidean theory, the three point function is determined
up to $j_i$-dependent normalization. However, in the Lorentzian
case, the normalization may be changed when $x_{ij}$ crosses
the zero. Nevertheless, for the three point functions under
consideration, we were able to further restrict their functional
form into only four independent ones with the label 
$\omega, \bar \omega = 0,1/2$.

In the following, we determine the coefficients $C_{\omega,\bar \omega}$
which cannot be determined only from the group theoretic consideration.
For the purpose, we again compute the Fourier transform of the above
quantities, and compare with overlaps of the three wave functions given
in section \ref{momentum}. 
In the latter basis, we can perform the integral more easily because it
is merely the integral of three Bessel functions, however we cannot see
how much the $SL(2,\br)$ symmetry determines the functional form of
three point function.

Here we should remark that we can perform the Fourier transform only if
$ {\rm Re} \, \tilde j < 2,\, {\rm Re} \, \tilde j_i > 0  \, (i=1,2,3) $.
For $j_i = 1/2 + i \omega_i$ $(i=1,2,3)$, this condition is always
satisfied, but if $j_i >1/2$ for some $i=1,2,3$, then the condition may
be violated. If the condition is violated, then we use the analytic
continuation on $j_i$ as the coefficients $C_{\omega,\bar \omega} (j_i)$.
For simplicity, we only consider the case with $j=1/2+i\omega$. The
case with $j>1/2$ can be obtained by setting $\omega = - i(j-1/2)$ and
using the different normalization (\ref{WDJp}). 
First, we compare in the case with
$\lambda_i \mu_i > 0$ $(i=1,2,3)$ to fix the coefficients 
$C_{\omega, \bar \omega}(j_i)$. The Fourier transform of the primary field in
this case does not depend on the choice of $A_j$, and we can fix the
coefficients completely. Then, we compare in the case with 
$\lambda_i \mu_i < 0$ $(i=1,2,3)$ as a consistency check.

\subsubsection{$\lambda_i \mu_i >0$ $(i=1,2,3)$ case}

Let us first see the overlaps of the three wave functions of the type
(\ref{WK}). Overlap integral of three Bessel functions is computable by
utilizing known results, for example, the appendix A of
\cite{ST1}. Making use of integral representations of the
hypergeometric functions collected in appendix \ref{formulae2}, we
found the three point function is given by\footnote{We set
$\lambda_1 \mu_2 \geq \lambda_2 \mu_1 $ without loss of
generality.}
\begin{align}
 &\left\langle \prod_{i=1}^3 \Phi^{\lambda_i,\mu_i}_{\omega_i} \right\rangle
 =  \frac{\pi}{\sqrt2} \delta(\lambda_1+\lambda_2+\lambda_3)
             \delta(\mu_1+\mu_2+\mu_3)
  (\lambda_3 \mu_3)^{i \tilde \omega - \frac12}
 \Gamma\left(\frac12 + i \tilde \omega \right)
 \prod_{j=1}^3 \frac{\Gamma\left(\frac12 - i \tilde \omega_j \right)}
   {\Gamma\left( 2 i \omega_j \right)} \nonumber \\
 & \quad \times \left(
 \frac{\sin \pi \left(\frac12 + i \tilde \omega \right)
       \sin \pi \left(\frac12 + i \tilde \omega_2 \right)}
      {\sin \pi (2 i \omega_2)} I_1 \bar I_1 -
 \frac{\sin \pi \left(\frac12 + i \tilde \omega_1 \right)
       \sin \pi \left(\frac12 + i \tilde \omega_3 \right)}
      {\sin \pi (2 i \omega_2)} I_2 \bar I_2
\right) ~.
\label{3pt}
\end{align}
Here, we abbreviated combinations of $\omega$-quantum numbers as
\begin{eqnarray}
  \tilde \omega = \omega_1 + \omega_2 + \omega_3  \qquad \mbox{and} \qquad
  \tilde \omega_j = \tilde \omega - 2 \omega_j ~.  
\end{eqnarray}
We also shorthanded the integrals
\begin{eqnarray}
 I_{(1,2)} = I_{(1,2)}\left(-\frac12+i\tilde\omega_1,-\frac12+i\tilde\omega_2;
  -\frac12+i\tilde\omega_3;-\frac{\lambda_2}{\lambda_3} \right),
\end{eqnarray}
and the barred ones $\bar I_{(1,2)}$ by replacing
$-\lambda_2/\lambda_3$ with $- \mu_2/\mu_3$.

We already know that the Fourier transform of the wave function with
$A_j= \pm 1$ is given by $(\ref{WK})$ independent of the parity. 
Therefore, we can determine the coefficients 
$ C_{\omega,\bar \omega}(j_i) $ by comparing the Fourier transform of 
(\ref{3pfform}) with (\ref{3pt}). We find that
\begin{align}
\left\langle
\prod_{i=1}^3 \Psi_{(-1)^{2\epsilon_i}}^{j_i}(x_i,\bar x_i)
\right\rangle
  &= \frac{\pi}{16\sqrt2}\Gamma (\tilde j - 1)
  \prod_{i=1}^3 \frac{1}{
 \sin \pi( j_i + \epsilon_i) \Gamma(2j_i -1)
 \Gamma (1- \tilde j_i)}   \nonumber\\ & \qquad \qquad \times
   \left(\sum_{\omega = 0,1/2}
  \frac{ (-1)^{2\omega} \sin \pi (\frac12 \tilde j + \omega) }{
  \prod_{i=1}^3\sin \pi (\frac12 \tilde j_i + \eta_i)}
     K_{\omega} (x_i) \bar K_{\omega} (\bar x_i) \right)~.
\label{dp3pt}
\end{align}
We should remark that the coefficients $ C_{\omega,\bar \omega}(j_i) $
depend on the parity of the wave functions. Therefore, for the general
three point functions (\ref{lc3pt}), the coefficients  
$ C_{\omega,\bar \omega}(j_i) $ depend on the parameters $A_j$ of the
primary fields (\ref{wfxx}).

\subsubsection{$\lambda_i \mu_i < 0$ $(i=1,2,3)$ case}

The overlaps of the three wave functions of the type (\ref{WCJ2})  
was essentially obtained in \cite{ST1}. 
There, the wave functions of the type (\ref{WCJ2}) with only the
phases $\Theta^{(0)}_{\nu_0=1} = - 1$ and $\Theta^{(1)}_{\nu_0=1} = + 1$ 
were considered. These wave functions are the ones with the
definite parity with $\pm 1$.
Denote these two types of wave function as
$\Phi_{0,\omega}^{\lambda,\mu}$ and
$\Phi_{1/2,\omega}^{\lambda,\mu}$, respectively, where the label
$\epsilon (= 0,1/2)$ keeps track of the parity $(-1)^{2 \epsilon}$. 
As before, the general wave functions (\ref{WCJ2})
are expressible as a linear combination of the two wave functions
of definite parity as
\begin{equation}
 \Phi_{\nu_0,\omega}^{(n)\lambda , \mu} = P^{(n)0}_{\nu_0}
 \Phi_{0,\omega}^{\lambda,\mu} + P^{(n)1/2}_{\nu_0}
 \Phi_{1/2,\omega}^{\lambda,\mu} ~.
\label{lc}
\end{equation}
It follows immediately that the general three point functions
are expressed as
\begin{equation}
 \left\langle \prod_{i=1}^3 \Phi^{(n_i)\lambda_i,\mu_i}_{\nu_{0,i},\omega_i}
 \right\rangle
 =
    \sum_{\begin{subarray}{c}
     \epsilon_1,\epsilon_2=0,1/2\\
     \epsilon_3=\epsilon_1+\epsilon_2 \mod 1
          \end{subarray}}
    P^{(n_1)\epsilon_1}_{\nu_{0,1}}
    P^{(n_2)\epsilon_2}_{\nu_{0,2}}
    P^{(n_3)\epsilon_3}_{\nu_{0,3}}
  \left\langle \prod_{i=1}^3 \Phi^{\lambda_i,\mu_i}_{\epsilon_i,\omega_i}
   \right\rangle ~.
\label{J3pt}
\end{equation}
Define the following integrals
\begin{equation}
 C_{\omega} (\lambda_i)
  = \int d x_1 d x_2 d x_3  K_{\omega} (x_i)
     e^{i \lambda_1 x_1 + i \lambda_2 x_2 + i \lambda_3 x_3} ~,
\end{equation}
then the three point functions (\ref{J3pt}) are written in terms
of these integrals as \cite{ST1}%
\footnote{Without loss of generality, we set $\lambda_1, \lambda_2
> 0$ and $\lambda_3 < 0$.}
 \begin{align}
 \left\langle \prod_{i=1}^3
  \Phi^{\lambda_i,\mu_i}_{\epsilon_i,\omega_i} \right\rangle
 &= \frac{(-1)^{2\epsilon_3}}{64\sqrt2}
  \prod_{i=1}^3 \frac{|\lambda_i|^{2 i \omega_i} e^{\pi i \epsilon_i}}{
 \sin \pi( \frac12 + \omega_i + \epsilon_i) \Gamma(2 i \omega_i) }
 \left(\sum_{\omega=0,1/2} \left(C_{\omega}(\lambda_i)\right)^{*}
  C_{\omega}(\mu_i) \right) ~. 
\end{align}
One can check that these Fourier transforms reproduce
(\ref{dp3pt}). In this expression, the latter factor may be
determined purely from group theoretic analysis \cite{ST1}, and
the former factor comes from the normalization of wave functions
(\ref{WCJ}).


\section{Conclusion}
\label{conclusion}

In this paper, we investigated string dynamics on Lorentzian $AdS_3$
in the minisuperspace limit.
We constructed Hilbert space by the normalizable wave functions, and
found that the Hamiltonian is given by a differential operator.
In order to see the self-adjointness of the Hamiltonian, 
we have to also determine the domain on which the Hamiltonian acts. 
For $\lambda\mu > 0$ case there is unique self-adjoint Hamiltonian,
however for  $\lambda\mu < 0$ case there is four parameter family of
self-adjoint extensions, and we picked up a domain labeled by 
one parameter $\nu_0$.
The condition of the self-adjointness reduces to the boundary condition
of the wave functions as (\ref{BC}), which can be interpreted as the
condition of the probability conservation or the current conservation.

We have constructed the Hilbert space by the square integrable functions
with respect to the inner product (\ref{innerprod}), and because of the
inner product, the eigenfunctions of the type (\ref{DJ}) have only 
$j > 1/2$. It is known \cite{BF} that there are two types of
solutions to the Klein-Gordon equation (\ref{KG}) for $0 < j < 1$ with
the same Casimir invariance $c_2 = j(j-1)$; 
one is given by $J_{2j -1}$ as in (\ref{DJ}) and the other is 
$J_{1-2j}$. In order to include the both of the solutions to the Hilbert
space, we have to modify the inner product as suggested in \cite{MO1} 
(but we did not because the unmodified inner product (\ref{innerprod})
gives the two point function of the minisuperspace limit of the
WZNW model). 
The ambiguity is related to the fact that
the AdS space is non-globally hyperbolic,\footnote{
In the Hilbert space we used, wave functions vanish sufficiently fast
near the boundary $z \to \pm 0$, where the non-globally hyperbolic
property of AdS space is important. Therefore, the every ambiguities
dealed with in this paper are related to the existence of the horizons in
the Poincar\'e patches of Lorentzian $AdS_3$, and not related to the
non-globally hyperbolic property.} 
and one parameter family of
self-adjoint extension is constructed in both the Poincar\'e patch
\cite{BBMS} and the global patch \cite{IW}. 
The application of the ambiguity to the AdS/CFT correspondence is
discussed in \cite{KW}.

Although we mainly considered in the Poincar\'e coordinates, the similar
analysis can be done in the global coordinates. As mentioned in
\cite{Balas1}, the solution to the Klein-Gordon equation is unique if we
require it behaves well near the center and the boundary. 
In fact, by closely following the analysis in $SL(2,\bc)/SU(2)$ WZNW case
\cite{Teschner1}, we can show that there is only unique self-adjoint
extension of the Hamiltonian. 
The difference from the Euclidean case is that there
appears discrete spectrum in $j>1/2$, in addition to continuous spectrum
in $j=1/2 + i \br$, which also exists in the Euclidean case.

The choice of coordinate system must not change the physics, so
the wave functions defined in the whole Poincar\'e patches should be
equivalent to the wave functions in the global coordinates. 
In other words, we should determine how to connect wave functions
defined in each Poincar\'e patch at the horizon so that the wave
functions reproduce the ones defined in the whole spacetime described
by the global coordinates. In this context, we could say that we did it
by assigning the self-adjointness of the Hamiltonian.
For instance, now we have the discrete label $j=1/2(n+\nu_0+1)$ with 
$n=0,1,\cdots$ for the wave functions with $j > 1/2$ (\ref{WDJ}).
In the global coordinates, the conserved charges are related to the
eigenvalues $(m,\bar m)$ of $J_R^3$ and $J_L^3$. 
For the states with $j > 1/2$ the quantum number $j$ is related to 
$m$ as $j = m + n$ or $j= -m + n$ with $n \in \bz$.
For the single cover of $AdS_3$, the label $m$ takes $m \in \frac12 \bz$
due to the closed timelike curve, and hence $j$ is also half integer.
For the universal cover, we unwrap the closed timelike curve,
and the label $m$ is replaced by $m = 1/2 (n + \nu)$ with 
$0 < \nu \leq 1$, where $\nu$ is integrated out later. 
Therefore, the label $j$ takes the same discrete value
$j=1/2(n+\nu_0+1)$.\footnote{I am grateful to Y.~Satoh for pointing out
this fact.}

In the Poincar\'e patch, there is no particular vacuum, and we have to
take care of the connection of the wave functions at the horizon.
The situation may be similar to the black hole case.
It is interesting to see if we can connect the wave functions between
horizons even in the case of black hole. Since the BTZ black hole can be
obtained by orbifolding the $AdS_3$ spacetime, we may be able to directly 
apply our analysis. 
If the black hole background is asymptotically AdS, then we
can apply our analysis to the AdS/CFT correspondence. 
Recently, in \cite{horizon,horizon1,horizon2,horizon3,horizon4}, it was
proposed that the information inside the horizon is obtainable from
the boundary CFT viewpoint. They use the analytic continuation, and our
analysis may be useful if we want to investigate in more general vacua.

Among the four parameter family of the self-adjoint extension, we pick
up a one parameter family in order to obtain the domain suitable to our
case. Apart from the choice used in the context, we may use a sensible
choice made in \cite{reduced1} and \cite{mini2}.\footnote{
The self-adjoint parameters are set $q_1 = q_2 = q_3 = 0$ in this case.}
Features of the choice are that there are no interaction between the
patches and that the wave functions given by the analytic continuation from
$\lambda \mu >0$ to $\lambda \mu < 0$ are not included in the domain of
self-adjoint Hamiltonian.
In \cite{mini2}, the timelike Liouville theory is defined using only a
patch. Thus, if we prepare two patches and glue
at the point where the expectation value of the tachyon diverges, then
the analytic continuation from the spacelike Liouville theory may work
as in \cite{S,GS2,MSY,ST,mini}. It would be interesting to pursue this
issue in more detail.

Moreover, we computed two and three point functions involving
primary fields. We used $(x,\bar x)$ representation since the 
$SL(2,\br)$ actions are expressible as differential operators 
(\ref{diffop}). Using the property under the $SL(2,\br)$ transformation,
the primary fields can be given as solutions to the differential
equations (\ref{diffeq}). The general solutions are given by
(\ref{wfxx}), which has one parameter $A_j$ in addition to the overall
normalization $C_j$. As we can see in (\ref{Fourier>}), there are maps
between $A_j$, $C_j$ and $a_{3,4}$ in (\ref{CJ}). In the Euclidean
theory, the solution to (\ref{diffeq}) is unique up to normalization
$C_j$ and the appearance of another parameter $A_j$ is a new feature
in the Lorentzian case. Functional forms of two and three point
functions are also fixed by the $SL(2,\br)$ symmetry as (\ref{2pfform})
and (\ref{3pfform}).
These solutions have several undetermined coefficients contrast to the
fact that there is only one undetermined coefficient in the Euclidean
case. Since the correlation functions in the minisuperspace model are
given by overlaps of the wave functions, we can compute them in the
minisuperspace approximation as (\ref{final}) and (\ref{dp3pt}).
Correlation functions in the full CFT may be computable if we use
the bootstrap constraint as in the Euclidean case \cite{Teschner2} as
well as the forms of the solutions obtained in this paper.


\subsection*{Acknowledgement}

I would like to thank Y.~Nakayama, Y.~Sugawara, T.~Takayanagi and
S.~Yamaguchi for useful discussions, and specially S.-J.~Rey for
important comments. I am also grateful to Harvard
University, University of Tokyo and Riken for the hospitality, where a
part of this work was done.

\appendix

\section{Several Useful Formulae}
\label{formulae}

\subsection{Integrals and related formulae}

Useful integrals reducing to the Gamma function are
\begin{align}
 &\int^a_0 dx (a^2 - x^2)^b = a^{1+2b} \sqrt{\pi}
 \frac{\Gamma(1+b)}{2\Gamma(\frac{3}{2}+b)} \qquad [a>0, \re b > -1] ~,\\
 &\int^{\infty}_a dx (x^2 - a^2)^b = a^{1+2b}
 \frac{\Gamma(-\frac12 - b) \Gamma(1+b)}{2\sqrt\pi} \qquad
   [a>0, -1 < \re b < -\frac12] ~,\\
 &\int^{\infty}_0 dx (a^2 + x^2)^b = a^{1+2b} \sqrt{\pi}
 \frac{\Gamma(-b-\frac12)}{2\Gamma(-b)} \qquad [\re b < -\frac12] ~,\\
 &\int^{\infty}_0 dx \frac{x^{a-1}}{(x+1)^{a+b}} =
 \frac{\Gamma(a) \Gamma(b)}{\Gamma(a+b)} \qquad [a,b>0] ~,\\
 &\int^{\infty}_1 dx x^{a-1}(x-1)^{b-1} =
 \frac{\Gamma(1-a-b) \Gamma(b)}{\Gamma(1-a)} \qquad [a+b < 1 , b>0] ~.
\end{align}
We use the following relations of the Gamma matrix as
\begin{equation}
 \Gamma (z) \Gamma (1-z) = \frac{\pi}{\sin \pi z} ~, \qquad
 \Gamma (2z) = \frac{2^{2z}}{2 \sqrt\pi}
  \Gamma(z) \Gamma\left(z+\frac12\right) ~.
\end{equation}
Fourier transforms used in the context are
\begin{align}
 &\int^{\infty}_{-\infty} dx \frac{1}{|x|^{\nu}} e^{-ixy}
    = 2 \sin \left(\frac{\nu\pi}{2}\right) \frac{\Gamma(1-\nu)}{|y|^{1-\nu}}
 \qquad [0 < \re \nu < 1] ~,\\
 &\int^{\infty}_{-\infty} dx \frac{{\rm sgn} x}{|x|^{\nu}} e^{-ixy}
    = 2 i {\rm sgn} y \cos \left(\frac{\nu\pi}{2}\right)
  \frac{\Gamma(1-\nu)}{|y|^{1-\nu}}
 \qquad [0 < \re \nu < 2 , \nu \neq 1] ~,\\
 &\int^{\infty}_{-\infty} dx \frac{1}{(x^2 + a^2)^{\nu+\frac12}} e^{-ixy}
    = \frac{2\sqrt\pi}{\Gamma(\nu+\frac12)}\left|\frac{y}{2a}\right|^\nu
      K_{\nu} (a|y|)
 \qquad [ \re \nu > -\frac12] ~,\\
 &\int_{|x|<a} dx \frac{1}{( a^2 - x^2)^{\nu+\frac12}} e^{-ixy}
    =  \sqrt\pi \Gamma \left(\frac12 - \nu \right)
   \left|\frac{y}{2a}\right|^\nu J_{-\nu} (a|y|)
 \qquad [ \re \nu < \frac12] ~,\\
 &\int_{|x|>a} dx \frac{1}{( x^2 - a^2)^{\nu+\frac12}} e^{-ixy}
    =  - \sqrt\pi \Gamma\left(\frac12 - \nu \right)
 \left|\frac{y}{2a}\right|^\nu N_{\nu} (a|y|) \nonumber \\
 &\qquad \qquad \qquad \qquad[ -\frac12 < \re \nu < \frac12]\qquad
  N_{\nu} (z) = \frac{1}{\sin \nu \pi}[\cos \nu \pi J_{\nu} (z) -
  J_{-\nu} (z)] ~.
\end{align}
The asymptotic forms of Bessel functions are
\begin{equation}
 J_{\nu} (z) \sim \sqrt{\frac{2}{\pi z}} 
 \cos\left(z-\frac12\pi(\nu+\frac12)\right)
 ~ (z \sim \infty) ~,\quad
 J_{\nu} (z) \sim \frac{1}{\Gamma(\nu+1)} \left(\frac{z}{2}\right)^{\nu}
 (z \sim 0) ~.
\end{equation}
The asymptotic expansion of modified Bessel function for large $|z|$ is
\begin{equation}
  K_{\nu} (z) \sim \sqrt{\frac{\pi}{2z}}e^{-z} \sum_{n=0}^{\infty}
 \frac{\Gamma(\nu + n + \frac12)}{n! \ \Gamma(\nu - n + \frac12)(2z)^{2n}} ~.
\end{equation}

\subsection{Hypergeometric functions}
\label{formulae2}

We use the following integrals as $(0<z<1)$
\begin{align}
 I_1(a,b,c;z)&=\int^{\infty}_1 dw w^a(w-1)^b(w-z)^c 
    \quad [ \re \ (-a-c) > \re \ (-a-b-c-1) > 0] \nonumber \\
             &=\frac{\Gamma(-a-b-c-1)\Gamma(b+1)}{\Gamma(-a-c)}
              F(-c,-a-b-c-1;-a-c;z) ~,\\
 I_2(a,b,c;z)&=\int^{z}_0 dw w^a (1-w)^b(z-w)^c 
    \quad [ \re \ (a+c+2) > \re \ (a+1) > 0] \nonumber \\
             &= z^{1+a+c}\frac{\Gamma(a+1)\Gamma(c+1)}{\Gamma(a+c+2)}
              F(-b,a+1;a+c+2;z) ~,
\end{align}
where $F(a,b;c;z)$ is the hypergeometric function.
Using a formula for the hypergeometric function
\begin{align}
 \frac{\Gamma(a)\Gamma(b)}{\Gamma(c)} F(a,b;c;z)
 &= \Gamma(a+b-c)(1-z)^{c-a-b}F(c-a,c-b;c-a-b+1;1-z) \nonumber \\
 & \qquad + \frac{\Gamma(a)\Gamma(b)\Gamma(c-a-b)}{\Gamma(c-a)\Gamma(c-b)}
    F(a,b;a+b-c+1;1-z) ~,
\end{align}
we can show that the above integrals satisfy the following relations
\begin{align}
 I_1(a,b,c;z)&=\frac{\sin \pi a}{\sin \pi (b+c)} I_1 (b,a,c;1-z)
              -\frac{\sin \pi c}{\sin \pi (b+c)} I_2 (b,a,c;1-z) ~,\\
 I_2(a,b,c;z)&=-\frac{\sin\pi(a+b+c)}{\sin \pi (b+c)} I_1 (b,a,c;1-z)
              -\frac{\sin \pi b}{\sin \pi (b+c)} I_2 (b,a,c;1-z) ~.
\end{align}
A formula of the hypergeometric function
\begin{equation}
 F(a,b;c;z)=(1-z)^{c-a-b} F(c-a,c-b;c;z)
\end{equation}
is also used.


\end{document}